\documentclass[%
 reprint,
superscriptaddress,
 amsmath,amssymb,
 aps,
prb,
]{revtex4-2}

\usepackage{graphicx}
\usepackage{dcolumn}
\usepackage{bm}
\usepackage{float}

\usepackage{hyperref}
\usepackage{soul}
\usepackage{afterpage}
\usepackage{xcolor}
\hypersetup{
    colorlinks=true,
    linkcolor={red!50!black},
    citecolor={blue!50!black},
    urlcolor={blue!80!black}
}
\renewcommand{\vec}[1]{\boldsymbol{#1}}
\newcommand{\hatb}[1]{\boldsymbol{\hat{{#1}}}}

\graphicspath{ {images/} } 

\begin{document}

\preprint{APS/123-QED}

\title{Parametric resonance of spin waves in ferromagnetic nanowires \\ tuned by spin Hall torque}

\author{Liu Yang}
\affiliation{Department of Physics and Astronomy, University of California, Irvine, California 92697, USA}%

\author{Alejandro A. Jara}%
\affiliation{Department of Physics and Astronomy, University of California, Irvine, California 92697, USA}%
\affiliation{Departamento de F\'isica, Universidad T\'ecnica Federico Santa Mar\'ia, Valpara\'iso 2390123, Chile}

\author{Zheng Duan}%
\author{Andrew Smith}%
\author{Brian Youngblood}%
\affiliation{Department of Physics and Astronomy, University of California, Irvine, California 92697, USA}%

\author{Rodrigo E. Arias}
\affiliation{Departamento de F\'{i}sica, FCFM, Universidad de Chile, Casilla 487-3, Santiago, Chile}

\author{Ilya N. Krivorotov}%
\affiliation{Department of Physics and Astronomy, University of California, Irvine, California 92697, USA}%

\date{\today}

\begin{abstract}
We present a joint experimental and theoretical study of parametric resonance of spin wave eigenmodes in Ni$_{80}$Fe$_{20}$/Pt bilayer nanowires. Using electrically detected magnetic resonance, we measure the spectrum of spin wave eigenmodes in transversely magnetized nanowires and study parametric excitation of these eigenmodes by a microwave magnetic field. We also develop an analytical theory of spin wave eigenmodes and their parametric excitation in the nanowire geometry that takes into account magnetic dilution at the nanowire edges. We measure tuning of the parametric resonance threshold by antidamping spin Hall torque from a direct current for the edge and bulk eigenmodes, which allows us to independently evaluate frequency, damping and ellipticity of the modes. We find good agreement between theory and experiment for parametric resonance of the bulk eigenmodes but significant discrepancies arise for the edge modes. The data reveals that ellipticity of the edge modes is significantly lower than expected, which can be attributed to strong modification of magnetism at the nanowire edges. Our work demonstrates that parametric resonance of spin wave eigenmodes is a sensitive probe of magnetic properties at edges of thin-film nanomagnets.

\end{abstract}

\pacs{Valid PACS appear here} 
\maketitle


\section{\label{sec:Introduction}Introduction}

Magnetization dynamics in thin-film nanoscale ferromagnets  is of fundamental and practical importance in the field of spintronics \cite{Hoffmann2015,Demidov2017, Parkin2016,Back2015,Hellman2017, Sander017, Sluka2019}. The spectrum of spin wave excitations in such nanomagnets is quantized due to geometric confinement \cite{Demokritov2002,Kostylev2005}, which gives rise to a plethora of interesting nonlinear magneto-dynamic effects not found in bulk ferromagnets \cite{Slavin2012,Tiberkevich2009,Ivanov2001,Grollier2017,Grollier2014,Cottam2017,Mancilla2016}. However, calculations of the spin wave spectrum in such structures are challenging due to the importance of nonlocal dipolar interactions \cite{Guslienko2002,Mills2005} and poor understanding of boundary conditions for dynamic magnetization at the nanomagnet edges \cite{Cowburn2000,McMichael2006,Krawczyk2008,Hammel2015}. Despite these challenges, a quantitative description of magnetization dynamics in nanomagnets is critically needed for design and optimization of nanoscale spintronic devices \cite{Fang2019, Tsunegi2019, Dieny2020, Talmelli2020} such as spin torque memory (STT-MRAM) \cite{Dorrance2013,Bhatti2017, Gajek2012b}, spin torque nano-oscillators \cite{Kiselev2003, Rippard2004, Grollier2014b,Ebels2007, Hache2020, Tarequzzaman2019, Koo2020} and ultrasensitive spintronic sensors \cite{Fuji2019}. Operation of all these practical spintronic devices critically depends on details of linear and nonlinear \cite{Li2019} magnetization dynamics in nanomagnets \cite{Barman2021}.

A significant body of prior experimental \cite{Kostylev2007, Katine2013,Almulhem2018,Baberschke2008,Chumak2020,Lindner2011,Gubbiotti2003, Livesey2013, Yu2016, Zhang2019, Purser2020, Schultheiss2021, Zhou2015, Liu2018, Barsukov2019} and theoretical \cite{Schmidt2012,McMichael2005,Mancilla2017,Krawczyk2014} work has been dedicated to studies of spin waves in nanostructures and their interactions with spin currents \cite{Slonczewski1996,Berger1996,Hirsch1999,Zhang2000,Finocchio2007,Ando2008,Dumas2013,Fan2013,Hoffmann2013,Bai2013, Padron-hernandez2011, Ganzhorn2016, Iacocca2019, Wang2019b, guckelhorn2020, Montoya2019}. These studies typically focus on the frequencies and spatial profiles of the eigenmodes. At present, a good quantitative understanding of many types of spin waves in  nanomagnets has been achieved with a notable exception of the eigenmodes localized near the nanomagnet edge, the so-called edge modes \cite{Silva2021,McMichael2013}. This is not surprising because magnetic properties of the edge of a thin magnetic film can differ from those of the rest of the film \cite{Silva2021,Silva2011,McMichael2007}, and also from sample to sample. Many magnetic properties such as magnetization, exchange interactions and magnetic anisotropy can become strongly spatially dependent near the magnetic film edge \cite{Belyaev2019,McMichael2012,McMichael2010}, and details of the magnetic edge profile are not well known \cite{McMichael2006b}. Measurements of the edge mode frequencies alone do not provide sufficient information to reconstruct the edge-induced modifications of the film magnetic properties. Therefore, characterization of the edge eigenmode properties going beyond the mode spectrum are needed. The relatively poor understanding of the edge eigenmodes is a challenging problem of significant practical importance because lateral dimensions of spintronic nanodevices such as STT-MRAM are projected to decrease down to a few nanometers \cite{Bhatti2017,Fullerton2008}, which implies that static and dynamic magnetic properties of such devices will be dominated by the magnetic film edge.

In this paper, we study spin wave eigenmodes in ferromagnetic thin-films nanowires \cite{Duan2014,Duan2014b,Duan2015, Yang2015, Smith2020} focusing on the edge eigenmodes \cite{Park2002}. The translational symmetry of the nanowire geometry significantly simplifies theoretical description of the spin wave spectrum and allows us to compare our measurements to an analytical theory of nanowire spin wave eigenmodes we develop here. In order to understand the eigenmode properties beyond the typically measured frequency and damping, we study parametric excitation of spin waves and its tuning by antidamping spin-orbit torque \cite{Miron2010,Ulrichs2011,Demidov2011,Rousseau2012,Wang2013,Hahn2013,Bracher2011,Guo2014,Urazhdin2010,Edwards2012,Epshtein2012, Garello2013, Geranton2016, Laczkowski2017, Ryu2019, Manchon2019, Wang2019, Belashchenko2020, Safranski2020, Shao2021, Kumar2021, Filianina2020, Bapna2018, Safranski2019}. To our knowledge, our experiment is the first measurement of parametric excitation of the edge spin wave eigenmodes. Measurements of the parametric resonance threshold and its tuning by antidamping spin Hall torque allows us to probe ellipticity of the edge modes. This new information on the properties of the edge modes allows us to test a popular model of the edge-induced modifications of thin film magnetic properties \cite{McMichael2006}. Our work places new constraints on the models of magnetic film edge and suggests a pathway for improving these models.

\section{\label{sec:setup}Samples and Measurements}

\begin{figure}[pt]
\includegraphics[width=0.9\columnwidth]{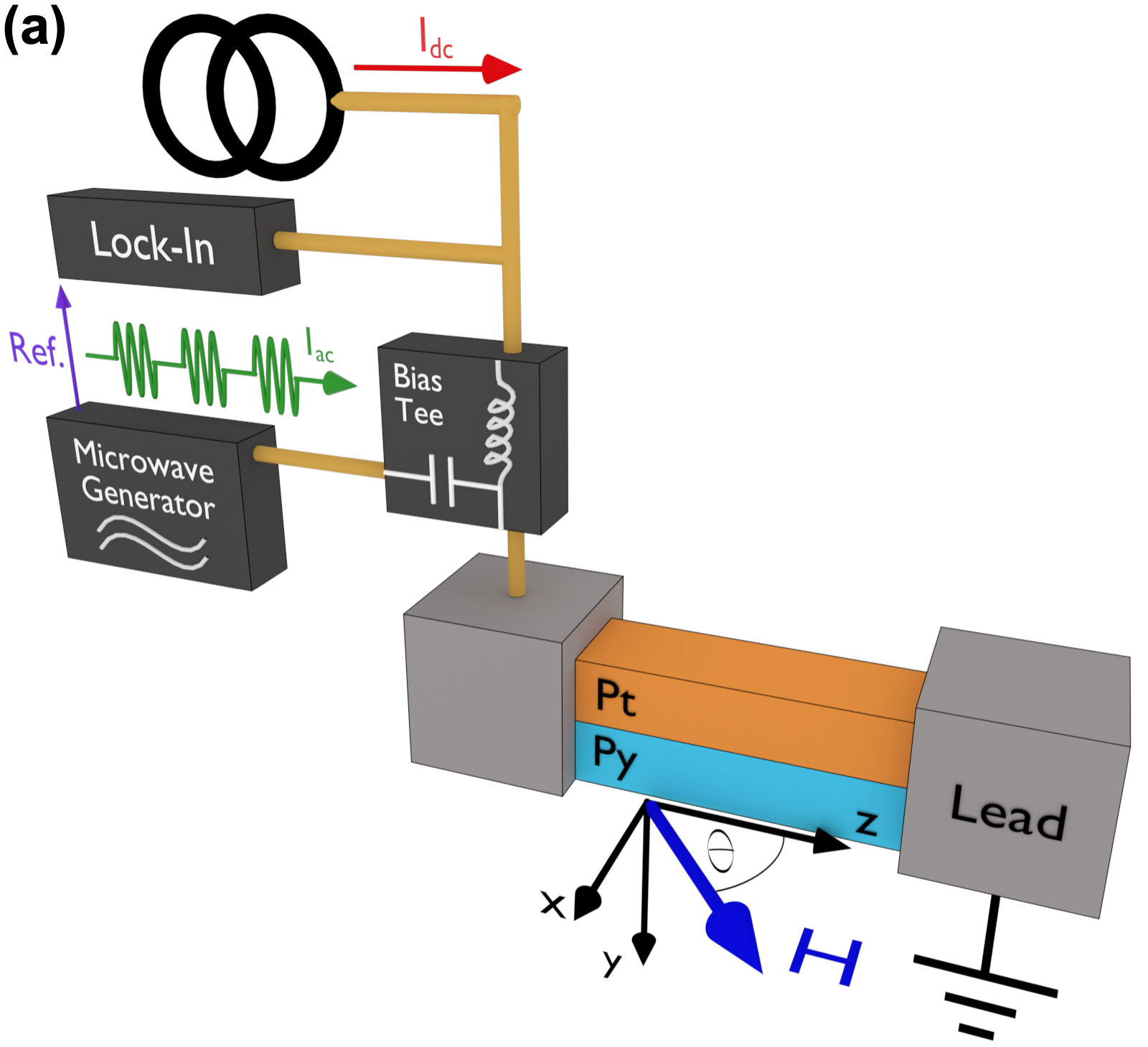}\\
\includegraphics[width=0.9\columnwidth]{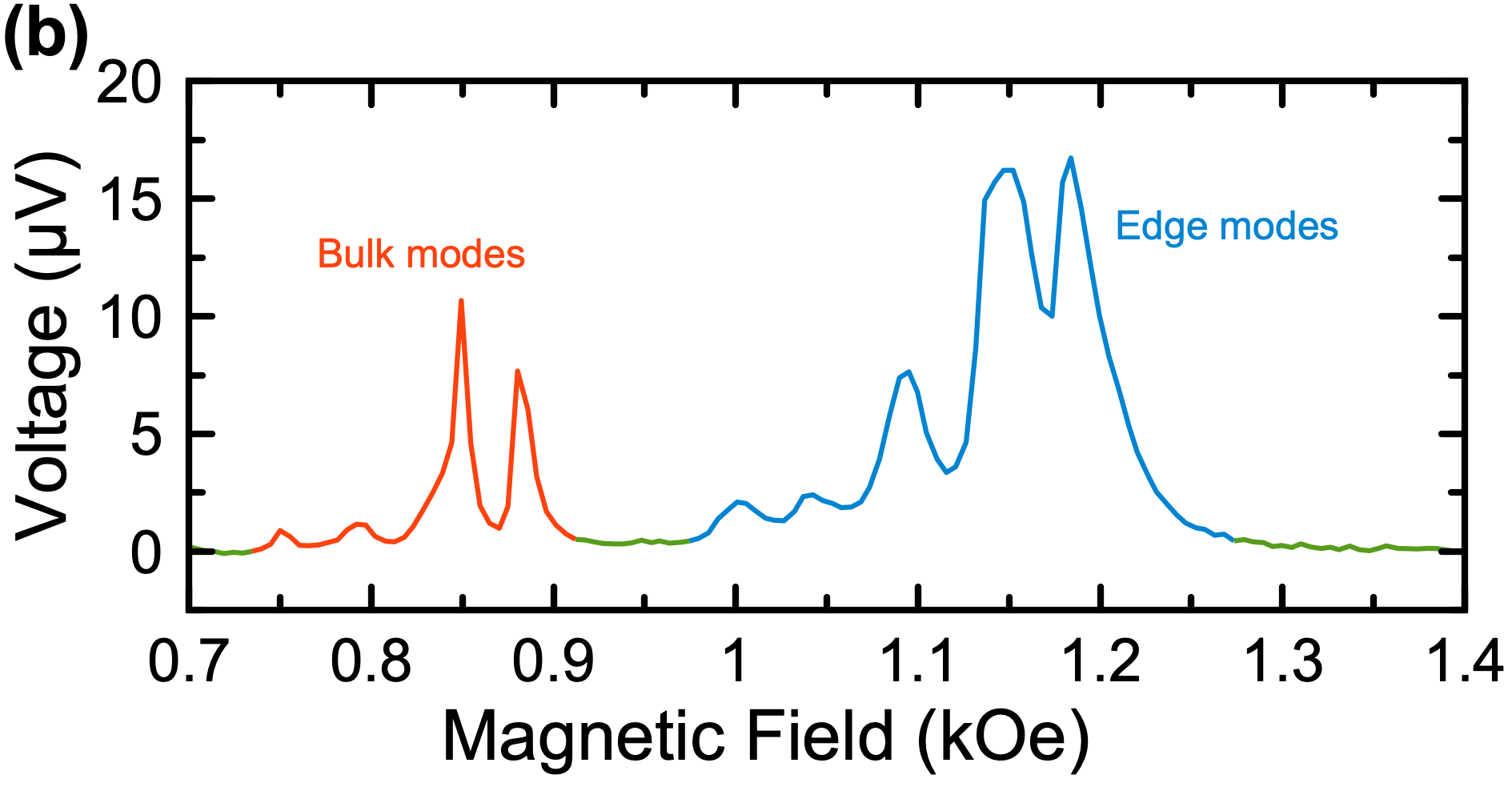}
\caption{\textbf{ST-FMR measurement schematic and an ST-FMR spectrum.} \textbf{(a)} ST-FMR measurement setup and the coordinate system used in this work. An amplitude-modulated microwave current $I_\mathrm{ac}$ from a microwave generator is applied to the Py/Pt nanowire device, and voltage $V$ induced at the modulation frequency is measured by a lock-in amplifier as a function of external field $H$ applied in the plane of the sample ($\it{xz}$-plane) at an angle $\theta$ with respect to the wire axis. A direct current  $I_\mathrm{dc}$ can be applied to the nanowire to tune its effective magnetic damping by spin Hall torque. \textbf{(b)} ST-FMR spectrum of the nanowire device measured at the microwave drive frequency of 6\,GHz, $\theta=85^{\circ}$, $I_{\mathrm{ac}}$ = 0.3\,mA, and $I_{\mathrm{dc}}$ = 2.2\,mA.}
\label{Fig:Layout}
\end{figure}

The nanowire devices studied in this work are patterned from GaAs(substrate)/AlO$_x$(4\,nm)/Py(5\,nm)/ Pt(5\,nm) multilayers deposited by magnetron sputtering, where Permalloy (Py) is a Ni$_{80}$Fe$_{20}$ alloy. Multilayer nanowires that are 6\,$\mu$m long and 190\,nm wide are defined via e-beam lithography and Ar plasma etching. Two Cr(7\,nm)/Au(35\,nm) leads are attached to each nanowire with a 1.8\,$\mu$m gap between the leads, which defines the active region of the device as shown in Fig.\,\ref{Fig:Layout}(a). 

We employ an electrically detected ferromagnetic resonance (FMR) technique also known as spin-torque FMR (ST-FMR) \cite{Tulapurkar2005,Sankey2006, Duan2014, Biziere2009, Ganguly2014, Goncalves2013, Cheng2013} to characterize spin waves in the nanowire. Figure~\ref{Fig:Layout}(a) shows the schematics of the ST-FMR setup, which allows us to measure both direct (linear) and parametric (nonlinear) excitation of spin waves in the Py nanowire. In these measurements, we apply an amplitude-modulated microwave current $I_\mathrm{ac}$ to the nanowire through the RF port of a bias tee, where $I_\mathrm{ac}$ represents the root mean square (rms) amplitude of the microwave current. This current applies periodic spin Hall torque and Oersted field $H_\mathrm{ac}$, both arising from microwave current in the Pt layer, to drive forced oscillations of the Py magnetization and thereby excite spin wave modes in the Py nanowire.  

We then measure voltage $V$ induced in the nanowire at the modulation frequency $f_\mathrm{mod}$ using a lock-in amplifier \cite{Sankey2006}. The measured voltage $V$ has two contributions \cite{Mecking2007}: (i) photovoltage signal arising from mixing of the microwave current $I_\mathrm{ac}$ and Py resistance oscillations $R_\mathrm{ac}$ at the microwave drive frequency $f$ and (ii) photoresistance signal arising from modulation of the time-averaged sample resistance at $f_\mathrm{mod}$ due to excitation of spin waves. Both the photovoltage and the photoresistance signals are due to anisotropic magneto-resistance (AMR) of the Py layer. As shown in Fig.\,\ref{Fig:Layout}(b), when $f$ coincides with the resonance frequency of a spin wave eigenmode, a peak is observed in the FMR spectrum $V(f)$ or $V(H)$. These measurements were made for magnetic field $H$ applied in the sample plane at the angle $\theta=85^{\circ}$ with respect to the electric current direction as illustrated in Fig.\,\ref{Fig:Layout}(a). Similar to the FMR spectra in our previous work \cite{Duan2014}, we observed two groups of modes: bulk and edge modes. These modes have different profiles along the wire width with reduced amplitude near the wire edges for the bulk modes, and enhanced amplitude for the edge modes. Several closely spaced bulk and edge modes are observed due to quantization induced by the geometric confinement of the modes along the wire length to the 1.8\,$\mu$m active region. Measurements in this work are performed at the bath temperature $T=4.2\,$K unless indicated otherwise. 

\begin{figure}[pt]
\includegraphics[width=0.99\columnwidth]{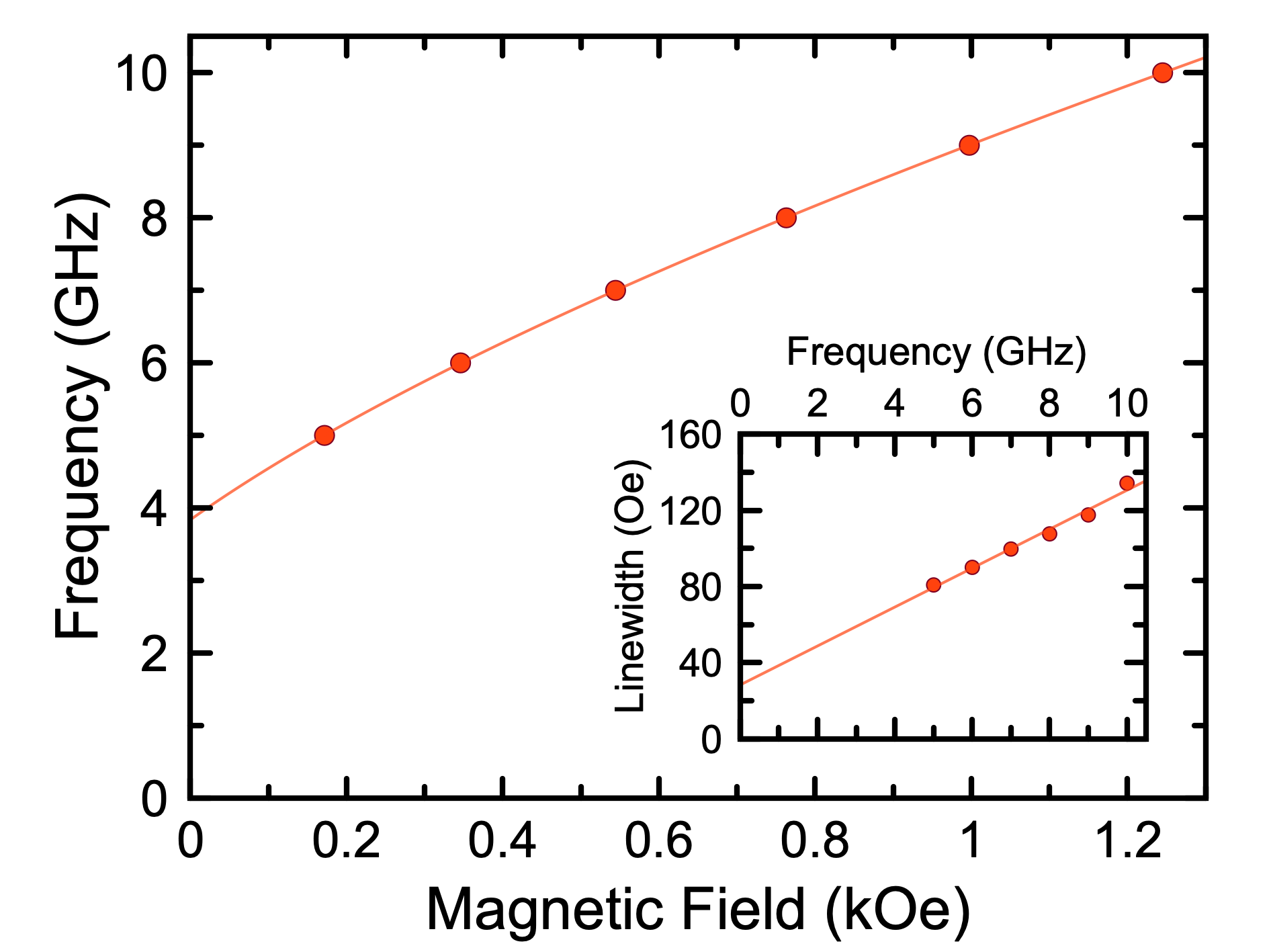}
\caption{\textbf{ST-FMR measurements at \boldmath{$\theta=0^\circ$, longitudinal magnetization}.} Resonance frequency of the quasi-uniform spin wave mode versus magnetic field applied parallel to the nanowire axis at the bath temperature $T=94\,$K. Inset shows linewidth of the mode versus frequency. Circles are experimental data while lines are fits described in the text.}
\label{Fig:FMR}
\end{figure}

In order to measure the Gilbert damping parameter of the nanowire, we apply an external magnetic field along the nanowire axis [$\theta=0^{\circ}$ in Fig.\,\ref{Fig:Layout}(a)] and measure resonance frequency and linewidth (half-width at half maximum) of the lowest-frequency (quasi-uniform) bulk mode, as shown in Fig.\,\ref{Fig:FMR}. The slope of the linewidth versus frequency in the inset of Fig.\,\ref{Fig:FMR} gives the effective damping of the quasi-uniform (bulk) mode: $\alpha = 0.031$, a value exceeding that of a thin Py film. This relatively high value of the damping parameter likely arises from two factors: (i) spin pumping into the proximate Pt layer and (ii) atomic inter-diffusion between Py and adjacent layers induced by heating in the device nanofabrication process. The measurements in Fig.\,\ref{Fig:FMR} were made at $I_{\mathrm{dc}}=0$ and $T=94$\,K  -- the temperature the wire reaches due to ohmic heating at bath temperature $T=4$\,K and $I_{\mathrm{dc}}$ = 2.2\,mA in Fig.\,\ref{Fig:Layout}(a).

To study parametric excitation of spin waves in the nanowire and tuning of this process by spin Hall current, we apply a  magnetic field $H>$ 450\,Oe in the plane of the sample at the direction perpendicular to the nanowire axis ($\theta=90^{\circ}\pm 0.1^{\circ}$). This field saturates Py magnetization perpendicular to the wire axis everywhere except very near the wire edges where demagnetizing field is enhanced by the edge magnetic charges \cite{Duan2015}. In this configuration, polarization of spin Hall current from Pt is nearly  parallel to  magnetization of Py, and modification of the effective damping of Py by spin Hall current is maximized \cite{Duan2014}. We apply a direct current $I_\mathrm{dc}$ to the nanowire in order to tune the effective damping of spin wave modes in Py by spin Hall torque arising from current in the Pt layer. In this paper, we use $I_\mathrm{dc}$  smaller than the critical current $I_\mathrm{c}$ for excitation of magnetization auto-oscillations by antidamping spin Hall torque \cite{Duan2014b}. 

For magnetization nearly saturated in the plane of the sample perpendicular to the nanowire axis, spin current polarization and the Oersted field are both parallel to magnetization and thus both spin torque and Oersted field torque are nearly zero. Therefore, direct excitation of spin waves by $I_\mathrm{ac}$ in this configuration is very inefficient. In addition, oscillations of magnetization at the ac current frequency $f$ give rise to resistance oscillations at $2f$ in this configuration due to the $R=R_0+R_A\cos^2\varphi$ angular dependence of AMR, with $\varphi$ the angle between magnetization and electric current. Therefore, mixing of resistance and current oscillation does not generate a rectified photovoltage (see Appendix \ref{STFMR} for details). Thus spin waves are both difficult to excite and detect electrically via application of $I_\mathrm{ac}$ at the spin wave resonance frequency for a magnetic field applied at $\theta=90^{\circ}$.

In contrast, $\theta=90^{\circ}$ is the optimum field direction for parametric excitation of spin waves in the nanowire. For efficient parametric excitation, either the external magnetic field parallel to the equilibrium magnetization direction or the effective damping of a spin wave mode (or both) should be modulated at twice the mode resonance frequency \cite{Gurevich1996}. For $\theta=90^{\circ}$, both the component of the Oersted field parallel to magnetization and the modulation of the effective damping by spin Hall current from Pt are maximized. Therefore, application of $I_\mathrm{ac}$ at $2f$ can efficiently excite parametric resonance of spin waves in the Py nanowire for $\theta=90^{\circ}$. At the same time, spin wave excitations generate resistance oscillations at $2f$, which mix with $I_\mathrm{ac}$ at $2f$ to produce a non-zero rectified photovoltage. Therefore, the efficiency of parametric excitation and electrical detection of spin waves is maximized at $\theta=90^{\circ}$. Parametric excitation is a threshold effect and thus $I_\mathrm{ac}$ exceeding a threshold value $I_\mathrm{th}$ is required for excitation of spin waves at zero temperature. At a finite temperature, parametric drive amplifies the amplitude of thermal spin waves below the threshold current. Analytical expressions for the dependence of $V$ on the drive current $I_\mathrm{ac}$ are derived in Appendix \ref{STFMR} for the $I_\mathrm{ac}\gg I_\mathrm{th}$ and $I_\mathrm{ac}\ll I_\mathrm{th}$ limits: 

\begin{figure}[pt]
\includegraphics[width=0.99\columnwidth]{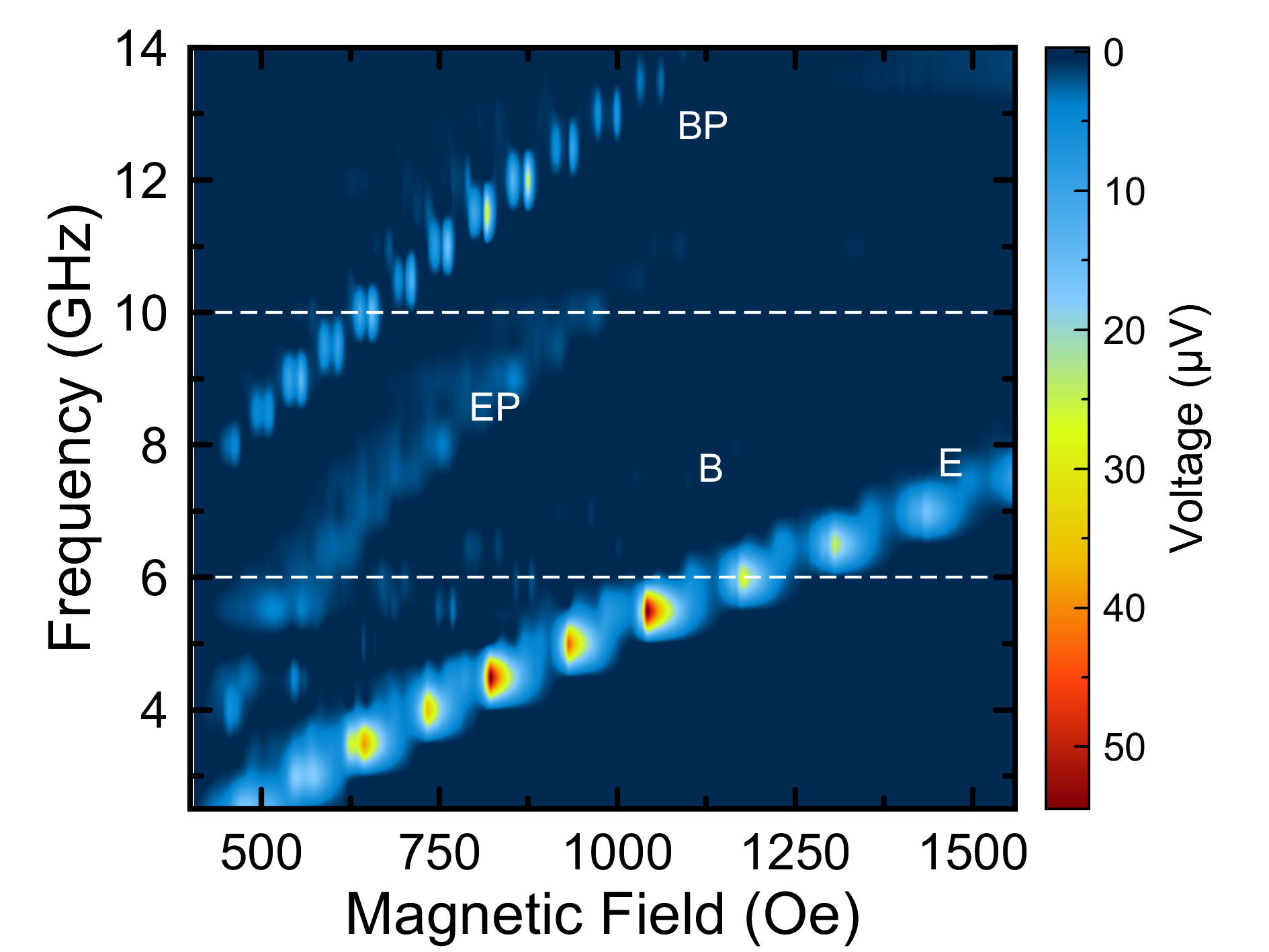}
\caption{\textbf{ST-FMR spectra at \boldmath{$\theta=90^\circ$}, transverse magnetization.} ST-FMR signal $V(f,H)$ measured as a function of ac frequency and magnetic field applied at $\theta=90^{\circ}$ for $I_{\mathrm{ac}}$ = 0.3\,mA and $I_{\mathrm{dc}}$ = 2.2\,mA. E, B, EP, and BP label directly excited edge mode, directly excited bulk mode, parametrically driven edge mode, and parametrically driven bulk mode, respectively. Dotted white lines highlight data at 6\,GHz and 10\,GHz employed for detailed analysis described in the text.}
\label{Fig:VoltageSpectra}
\end{figure}

\begin{eqnarray}
      V \sim 
    \begin{cases}
      \left(2I_\mathrm{dc}+\sqrt{2}I_\mathrm{ac}\right)/(I_\mathrm{th}-I_\mathrm{ac})^2 & \text{$I_\mathrm{ac}\ll I_\mathrm{th}$ }\\
      \left(2I_\mathrm{dc}+\sqrt{2}I_\mathrm{ac}\right)\sqrt{I_\mathrm{ac}^2-I_\mathrm{th}^2} & \text{$I_\mathrm{ac}\gg I_\mathrm{th}$. }
    \end{cases}
    \label{eq:parametric}
\end{eqnarray}

\begin{figure*}[pt]
\includegraphics[width=0.8\columnwidth]{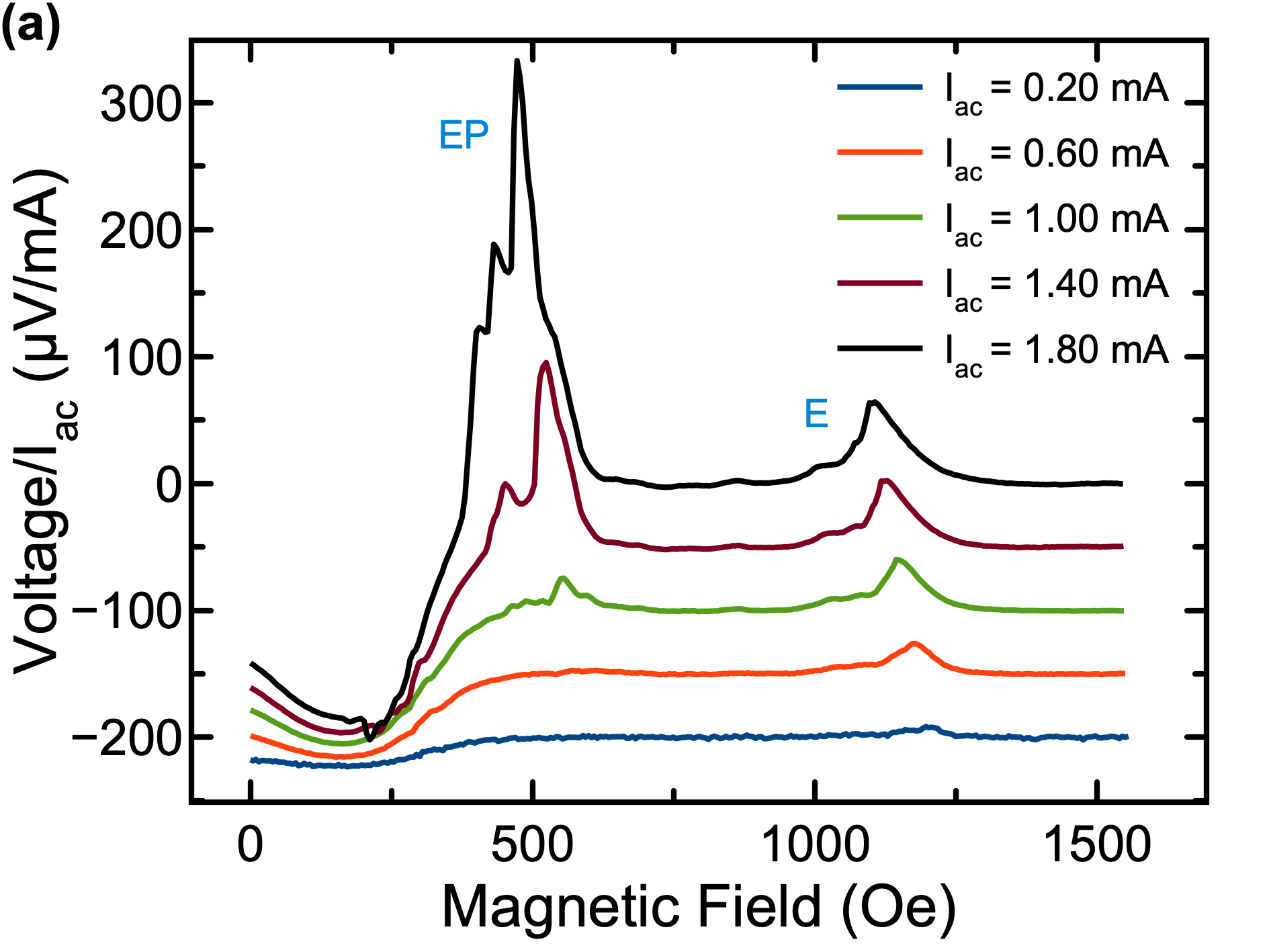}
\includegraphics[width=0.6\columnwidth]{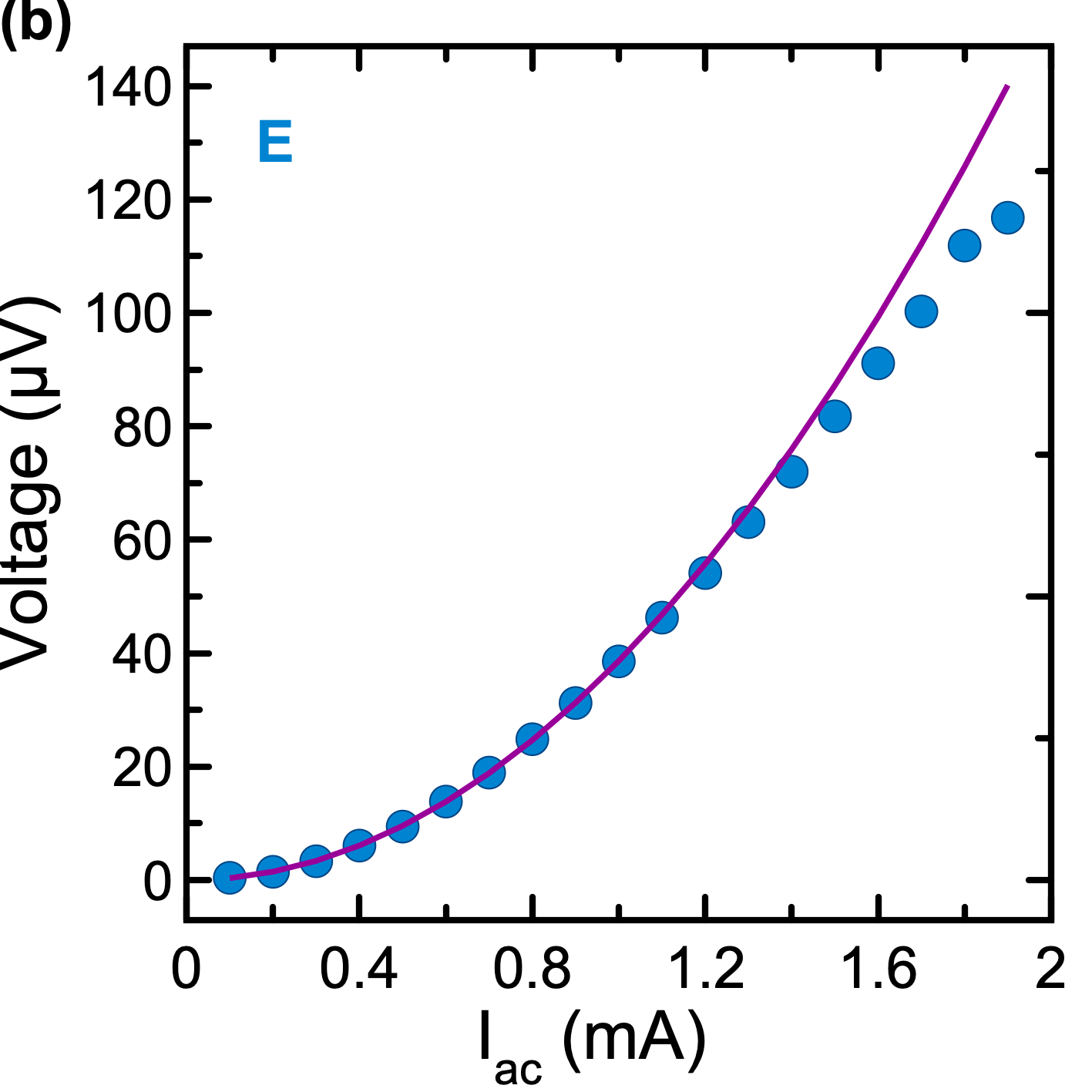}
\includegraphics[width=0.6\columnwidth]{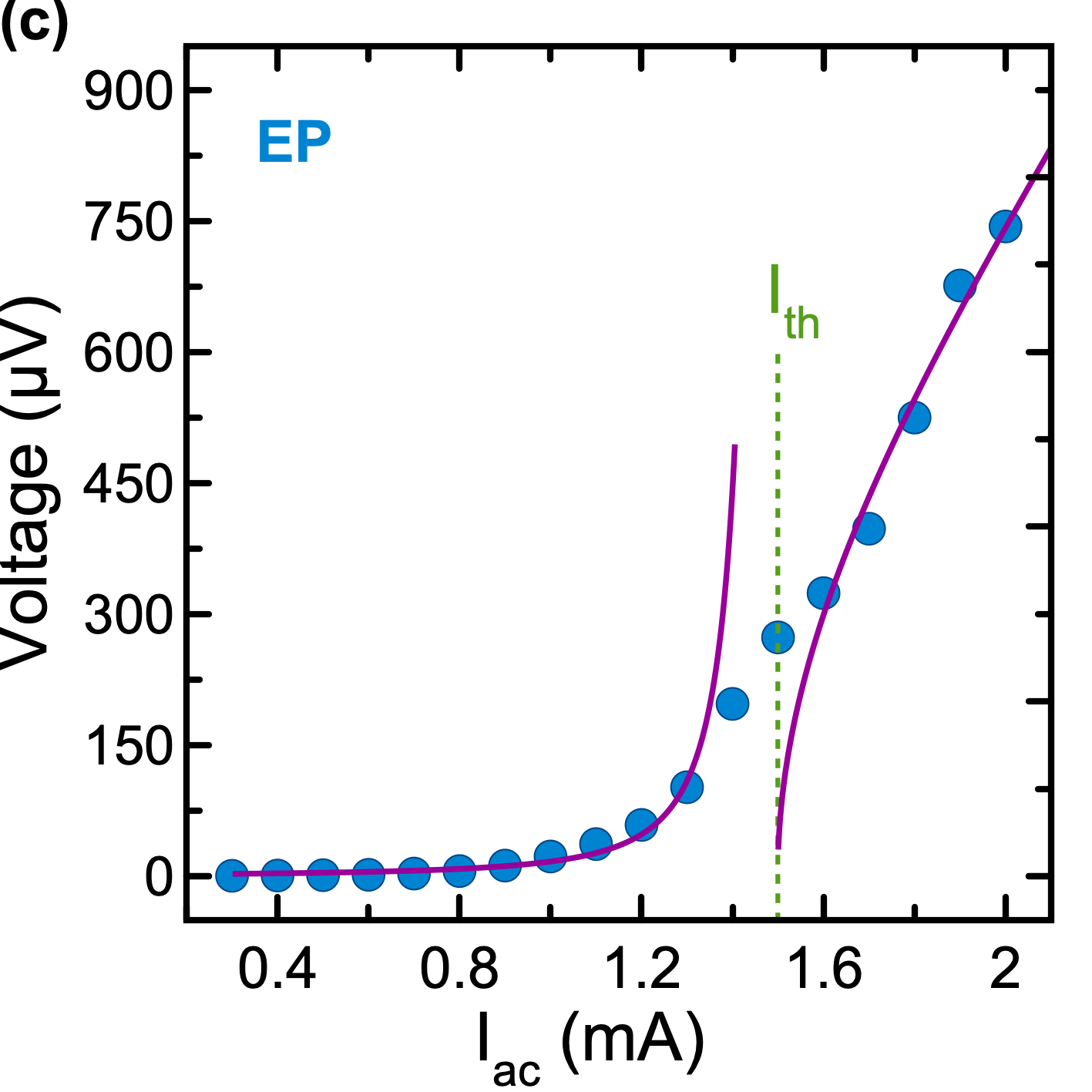}
\caption{\textbf{Dependence of ST-FMR spectra on ac current.} (a) ST-FMR spectra measured at five values of $I_\mathrm{ac}$, f\,=\,6\,GHz and $I_\mathrm{dc}$ = 1.8\,mA (vertically offset for clarity). (b) Directly excited edge mode amplitude as a function of $I_\mathrm{ac}$ and (c)  parametrically excited edge mode amplitude as a function of $I_\mathrm{ac}$. Lines are fits to Eq.\,(\ref{eq:direct}) and Eq.\,(\ref{eq:parametric}), respectively.}
\label{Fig:VoltageSignals}
\end{figure*}

Spin pumping combined with inverse SHE in the Pt layer can also give rise to an additional dc voltage term \cite{Bai2013,Liu2011}. However, due to its second order in spin Hall angle $\theta_\mathrm{SH}$, as well as the strong ellipticity of the oscillation, this contribution is orders of magnitude smaller than the signal given by Eq.\,(\ref{eq:parametric}) and is negligible \cite{Liu2011}.

\section{\label{sec:Results}Experimental Results and Analysis}

Figure~\ref{Fig:VoltageSpectra} shows ST-FMR spectra measured as a function of frequency and magnetic field applied at $\theta=90^{\circ}$, with $I_\mathrm{ac}$ = 0.3\,mA and $I_\mathrm{dc}$ = 2.2\,mA. This $I_\mathrm{dc}$ value is just below the critical current for the excitation of auto-oscillations of magnetization $I_\mathrm{c}$, which means that the effective damping is positive but close to zero. Multiple peaks are observed in the spectra. A comparison to ST-FMR data from a similar sample \cite{Duan2014b} lets us identify the two lowest frequency peaks as directly excited edge and bulk spin wave modes (marked as E and B, respectively) \cite{McMichael2006}. The bulk mode amplitude rapidly decreases with increasing hard-axis magnetic field $H$, as expected for a direct mode excitation by an ac drive parallel to magnetization. In contrast, ST-FMR signal amplitude of the directly excited edge mode is not small even for the largest field of 1.5\,kOe used in the measurement. The direct drive can efficiently excite the edge mode  because magnetization at the edge of the nanowire is not fully saturated along the applied field due to the high demagnetization field near the wire edges \cite{McMichael2006}. 

Two additional ST-FMR peaks are observed in Fig.\,\ref{Fig:VoltageSpectra} at frequencies close to twice the edge and bulk mode frequencies. These peaks marked as EP and BP arise from parametric excitation of the edge and bulk modes, respectively. 

Figure~\ref{Fig:VoltageSpectra} reveals that the parametrically excited bulk peak has a higher amplitude compared to the directly excited bulk peak due to the high efficiency of parametric excitation for magnetization parallel to the magnetic field. This trend is not observed for the edge mode because edge magnetization is not fully aligned with the applied field direction.

Figure~\ref{Fig:VoltageSignals}(a) illustrates the dependence of ST-FMR spectra on the amplitude of the drive $I_\mathrm{ac}$ at fixed dc current, $I_\mathrm{dc}$ = 1.8\,mA, and fixed frequency, 6\,GHz. Comparing to Fig.\,\ref{Fig:VoltageSpectra}, we identify the peak at 0.5\,kOe as the parametrically excited edge mode, and the peak at 1.1\,kOe as the directly excited edge mode. Figures~\ref{Fig:VoltageSignals}(b) and \ref{Fig:VoltageSignals}(c) show the magnitude of the peaks in Fig.\,\ref{Fig:VoltageSignals}(a) as a function of $I_\mathrm{ac}$. As expected \cite{Tiberkevich2009}, the magnitude of the ST-FMR peak for the directly excited edge mode increases quadratically with the amplitude of the eigenmode, which is proportional to $I_\mathrm{ac}$, as shown in Fig.\,\ref{Fig:VoltageSignals}(b):
\begin{eqnarray}
    V \propto I_\mathrm{ac}^{2}.
\label{eq:direct}
\end{eqnarray}

In contrast, the parametrically excited edge mode shows a threshold behavior in $I_\mathrm{ac}$ with rapid growth of the mode amplitude above a threshold drive value $I_\mathrm{th}$, as shown Fig.\,\ref{Fig:VoltageSignals}(c). We determine the value of $I_\mathrm{th}$ via fitting the data in Fig.\,\ref{Fig:VoltageSignals}(c) to Eq.\,(\ref{eq:parametric}). The best fit in this figure is shown by lines in both the $I_\mathrm{ac}\gg I_\mathrm{th}$ and $I_\mathrm{th}\gg I_\mathrm{ac}$ regimes with the common $I_\mathrm{th}$ fitting parameter.

Figure~\ref{Fig:VoltageSignals}(a) also shows that the linewidth of the ST-FMR peak increases with increasing amplitude for the parametrically excited mode. This increase happens via peak broadening towards lower resonance field (higher resonance frequency), which indicates that spin waves with shorter wavelength along the wire are excited at higher drive power. Specifically, Fig.\,\ref{Fig:VoltageSignals}(a) reveals a series of peaks that appear at lower resonance fields (higher frequencies) with increasing drive power. These peaks result from confinement of the edge mode to the active region along the wire length by the Oersted field \cite{Duan2014b}. The threshold current for parametric excitation of these higher frequency edge modes is higher for higher mode frequency due to smaller ellipticity of modes with shorter wavelengths \cite{Gurevich1996}. Indeed, at the highest ac current of excitation in Fig.\,\ref{Fig:VoltageSignals}(a) we observe two smaller side peaks at magnetic fields below the main peak at approximately at 500\,Oe. These side peaks arise from spin wave quantization along the wire length due to confinement to the $L_a=1.8\,\mu$m active region. We assume pinning of these modes at the ends of the active region due to the confining potential of the Oersted field from direct bias current in the Pt layer \cite{Duan2014, Duan2014b}. The pinning boundary conditions at the ends of the active region give longitudinal  wavelengths of the three lowest frequency modes of 3.6\,$\mu$m, 1.8\,$\mu$m and 1.2\,$\mu$m respectively. The magnetostatic Damon-Eshbach  character of these modes with wave vectors along the wire gives rise to a linear frequency-wavevector dispersion \cite{DamEsh1960}. Given this linear dispersion, we expect frequency-equidistant mode separation $\delta f$, which is broadly  consistent with the data in Fig.\,\ref{Fig:VoltageSignals}(a). Indeed, $\delta f$ in this case may be estimated from the low wavevector form of the magnetostatic  Damon-Eshbach frequencies of a film, i.e.  $f=G(\sqrt{h(h+1)} + (kb)/2\sqrt{h(h+1)})$, with $G=21.7$\,GHz  and $h=H/4\pi M_s$, where $M_s$ is saturation magnetization of our Py film. This gives a frequency separation between the neighboring length modes of $\delta f \approx$ 0.18\,GHz, which corresponds to a magnetic field separation between the length modes of $\delta H \approx$ 30\,Oe: see Appendix \ref{LM} for details. The   experimentally observed separation of these length modes  is approximately  $\delta H \approx$ 28 and 41\,Oe. Thus, given this quite close agreement, and the approximate nature of our theoretical explanation (the formula is valid for an infinite film, the effective magnetic field is lowered close to the edges of the stripe due to demagnetizing effects), we  may say that the physical explanation of these different peaks is quantization of modes along the longitudinal direction.

\begin{figure}[pt]
\includegraphics[width=0.99\columnwidth]{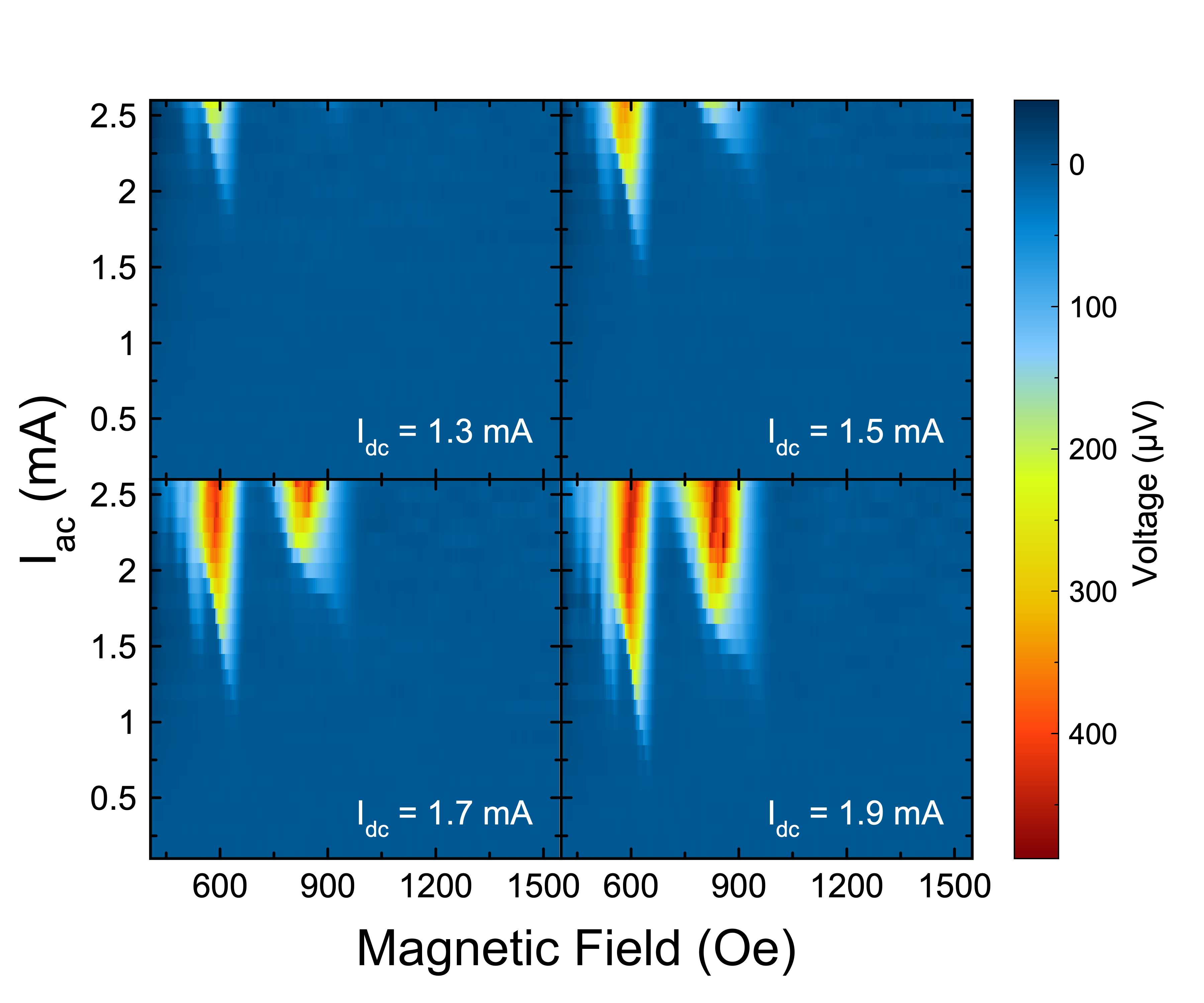}
\caption{\textbf{Effect of direct bias current on parametric excitation of bulk and edge modes.} ST-FMR signal measured at $f$\,=\,10\,GHz and four values of $I_{\mathrm{dc}}:$ \,1.3, 1.5, 1.7, and 2.2\,mA as a function of $I_\mathrm{ac}$ and $H$ applied at $\theta = 90^\circ$.}
\label{Fig:VoltageSpectra_4Idc_10GHz}
\end{figure}

Figure~\ref{Fig:VoltageSpectra_4Idc_10GHz} shows the dependence of ST-FMR signal on $I_\mathrm{ac}$ at 10\,GHz and $H$ applied at $\theta = 90^\circ$. Four panels of this figure show the data taken at four values of $I_\mathrm{dc}$. Parametrically excited bulk and edge mode signals are observed near 0.7\,kOe and 1\,kOe, respectively. This figure clearly illustrates the threshold character of the parametric spin wave excitation and shows the dependence of   $I_\mathrm{th}$ on the magnetic field.

\begin{figure*}
\includegraphics[width=0.65\columnwidth]{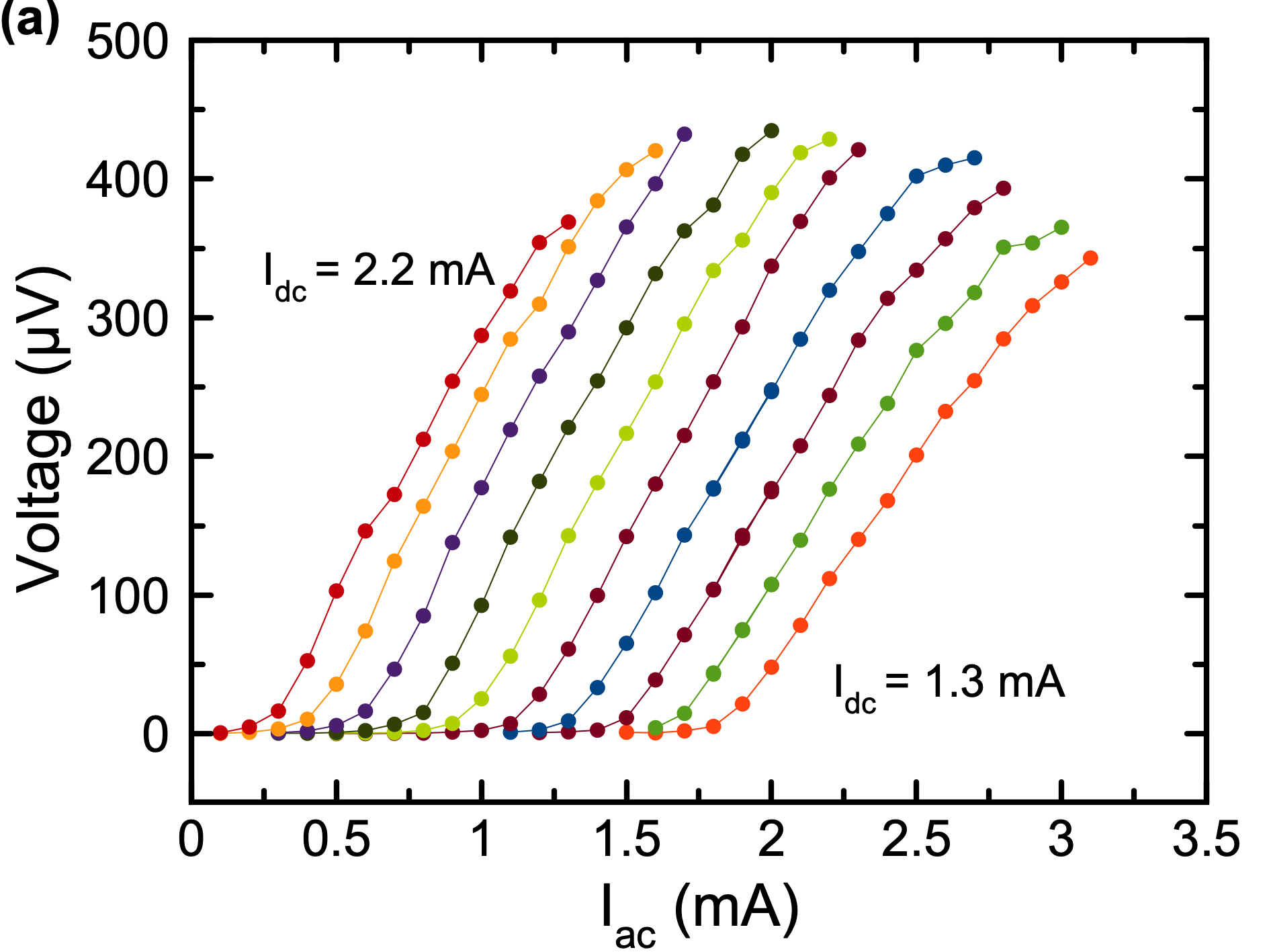}
\includegraphics[width=0.65\columnwidth]{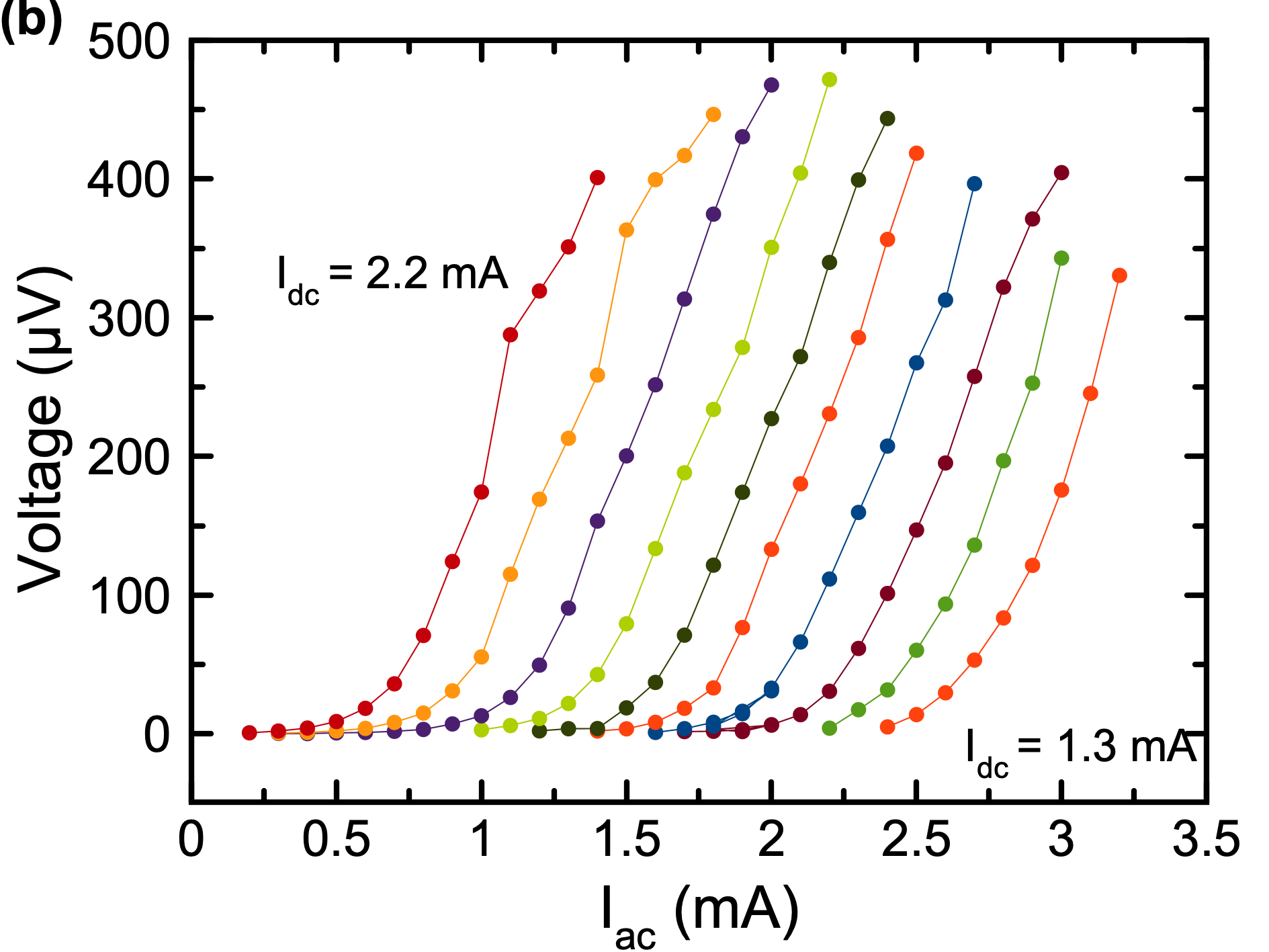}
\includegraphics[width=0.65\columnwidth]{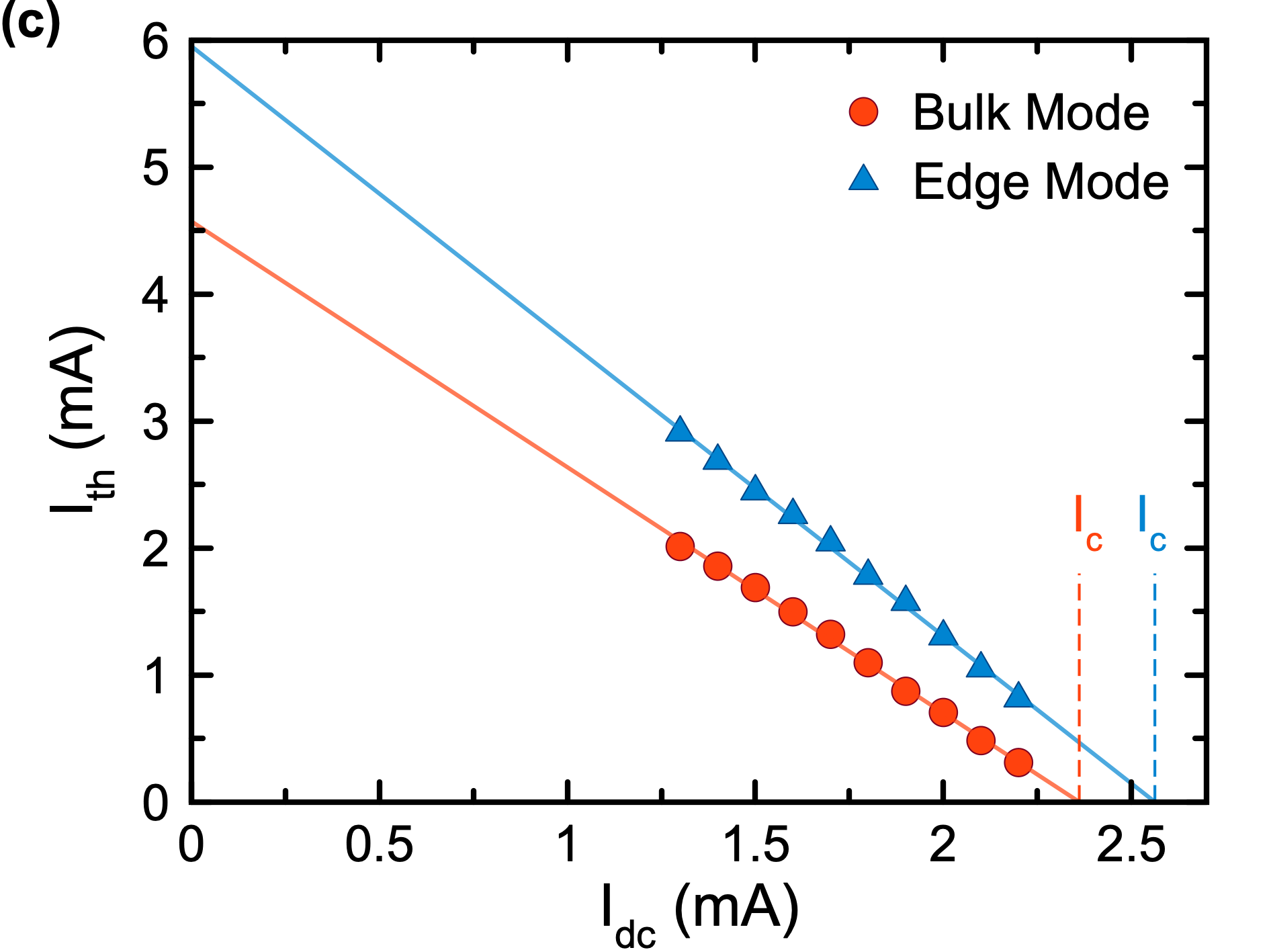}
\caption{\textbf{Tuning of parametric excitation by direct current}.
Parametric resonance peak voltage as a function of $I_\mathrm{ac}$ measured at 10\,GHz for $I_\mathrm{dc}$ ranging from 1.3\,mA to 2.2\,mA: (a) bulk mode and (b) edge mode. Lines are guides to the eye. (c) $I_\mathrm{th}$ as a function of $I_\mathrm{dc}$ measured for the bulk mode (circles) and edge mode (triangles). Lines are linear fits to the data.} 
\label{Fig:ParametricallyExcited_2}
\label{Fig:FittedRFthreshold}
\end{figure*}

Figure~\ref{Fig:VoltageSpectra_4Idc_10GHz} also reveals the effect of $I_\mathrm{dc}$ on $I_\mathrm{th}$. Antidamping spin Hall torque from $I_\mathrm{dc}$ decreases the effective damping of the modes with increasing $I_\mathrm{dc}$, which leads to a linear decrease of $I_\mathrm{th}$ with $I_\mathrm{dc}$. Figure~\ref{Fig:VoltageSpectra_4Idc_10GHz} also clearly shows that up to four bulk modes are excited parametrically. Similar to the case of the edge modes in Fig.\,\ref{Fig:VoltageSignals}(a), multiple bulk modes arise from spin wave confinement along the wire length within the active region of the nanowire. The threshold current for parametric excitation increases with increasing wavelength of the bulk mode along the wire length primarily due to decrease of the mode ellipticity with increasing wavelength \cite{Gurevich1996}.

Figures \ref{Fig:ParametricallyExcited_2}(a), \ref{Fig:ParametricallyExcited_2}(b) and \ref{Fig:ParametricallyExcited_2}(c) reveal further details of the dependence of $I_\mathrm{th}$ of the edge and bulk modes on $I_\mathrm{dc}$. Figures \ref{Fig:ParametricallyExcited_2}(a) and \ref{Fig:ParametricallyExcited_2}(b) show the ST-FMR peak amplitude for parametrically excited bulk and edge modes as a function of $I_\mathrm{ac}$ for different values of $I_\mathrm{dc}$. We fit each trace to Eq.\,(\ref{eq:parametric}) in order to extract quantitative values of $I_\mathrm{th}$ as a function of $I_\mathrm{dc}$. Symbols in Fig.\,\ref{Fig:ParametricallyExcited_2}(c) show $I_\mathrm{th}$ versus $I_\mathrm{dc}$ for the lowest-frequency bulk and edge modes obtained via this fitting procedure. The data in Fig.\,\ref{Fig:ParametricallyExcited_2}(c) reveal that $I_\mathrm{th}(I_\mathrm{dc})$ is a linear function with a negative slope, as expected due to the linear dependence of the effective damping on antidamping spin Hall torque.

A linear fit to the data in Fig.\,\ref{Fig:ParametricallyExcited_2}(c) allows us to precisely determine the critical current $I_\mathrm{c}$ for excitation of auto-oscillations of magnetization of the bulk and edge modes. This critical current is obtained as an intercept of the linear fit with abscissa of the plot. We note that this method of evaluation of $I_\mathrm{c}$  is significantly more precise than methods based on fitting of the microwave power emitted by the mode versus $I_\mathrm{dc}$ to theoretical values \cite{Akerman2014}, as is usually done for spin torque oscillators. This conventional method lacks precision due to thermally-activated excitation of the mode that smears out the auto-oscillation threshold and typically leads to under-estimation of $I_\mathrm{c}$. Thus our measurements of parametric excitation of spin wave modes demonstrate a precise method for measuring the threshold current for auto-oscillatory dynamics driven by anti-damping spin torques.

Extrapolation of the data in Fig.\,\ref{Fig:ParametricallyExcited_2}(c) to $I_\mathrm{dc}=0$ yields the values of $I_\mathrm{th}$ for the bulk and edge modes in the absence of spin Hall torque. The measured values of $I_\mathrm{th}$ for the bulk and edge modes allow us to test models of spin wave eigenmodes in the nanowire geometry. Indeed, in the parallel pumping geometry studied here ($H_\mathrm{ac}$ is parallel to the nanowire magnetization) \cite{Gurevich1996}, $I_\mathrm{th}$ is directly proportional to the mode damping and inversely proportional to the mode ellipticity \cite{Chen2017}. Thus $I_\mathrm{th}$ diverges for vanishing mode ellipticity. In contrast, $I_\mathrm{c}$, which is also directly proportional to the mode damping, decreases with decreasing mode ellipticity and remains finite for vanishing ellipticity \cite{Grollier2003,Chen2011}. Therefore, measurements of $I_\mathrm{th}$ and $I_\mathrm{c}$ for a given mode allow one to simultaneously determine both the mode ellipticity and the mode damping. This information puts stringent constraints on spin wave eigenmode models, and thus our measurements serve as sensitive tests of spin wave dynamics in the ferromagnetic nanowire geometry. As we show in subsequent sections, our measurements of $I_\mathrm{th}$ and $I_\mathrm{c}$ prove that the currently used model of bulk spin wave modes provides adequate description of the experiment while the edge mode models must be improved to quantitatively describe the experimentally observed edge eigenmodes.

\section{\label{sec:TheoreticalMethods}Theoretical Methods}

In this section, we derive an approximate theory of spin wave eigenmodes in the nanowire geometry and calculate the threshold drive values for parametric excitation of these modes. We consider the nanowire geometry shown Fig.\,\ref{Fig:Layout}(a), i.e. vertically stacked Py and Pt wires of rectangular cross section, each 5\,nm $=2b$ thick and 190\,nm $=2c$ wide. A cartesian coordinate system used in our calculations is shown in Fig.\,\ref{Fig:Layout}(a). An in-plane magnetic field $H$ is applied along the $\hatb{x}$ direction perpendicular to the nanowire axis, and ac and dc electric currents are applied in the $\hatb{z}$ direction along the wire axis.

Our theory takes into account magnetic dilution at the nanowire edges.  In this model first proposed in Ref.~\cite{McMichael2006}, the magnitude of the magnetization near the wire edge depends on the distance from the edge, $|\vec{M}(\vec{x})|=M_\mathrm{s}(x)$. Specifically, $M_\mathrm{s}(x)$ is assumed to grow linearly from zero at the edge to its maximum value $M_0$ (saturation magnetization) over the edge dilution length $L$ \cite{Duan2014b}. The model assumes that the exchange constant of the ferromagnet is proportional to $M_\mathrm{s}^2(x)$. Details of a theoretical treatment of the dilution region within a continuum model can be found in Ref.~\cite{Kruglyak2014}.

The dilution model was used in Ref.~\cite{Duan2014b} to fit experimentally measured in-plane and out-of-plane saturation fields as well as the bulk mode eigenfrequency for the Py/Pt nanowires studied here. This fitting procedure gave $M_0 = 608$\,emu/cm$^3$, $L=10$\,nm and $K_\mathrm{s} = 0.237$\, erg/cm$^2$, where $K_\mathrm{s}$ describes interfacial perpendicular magnetic anisotropy in this system.

We determine the spin wave dynamics in our nanowire system via solving the Landau-Lifshitz-Gilbert (LLG) equation:
\begin{widetext}
\begin{equation}
\frac{d\vec{M}}{dt} =-|\gamma| \vec{M} \times \vec{H}_\mathrm{eff}+|\gamma| 4 \pi J \vec{M} \times ( \vec{M} \times \hatb{x} ) +\alpha\frac{\vec{M}}{M_\mathrm{s}} \times \frac{d\vec{M}}{dt} \; .
\label{LLG}
\end{equation}
\end{widetext}
The first term in Eq.\,(\ref{LLG}) describes precession of the magnetization around an effective magnetic field $\vec{H}_\mathrm{eff}$, the second term describes spin Hall torque, and the third term describes magnetic damping parametrized by the Gilbert damping constant $\alpha$. We assume uniform magnetization over the 5\,nm thickness of Py because it is similar to the Py exchange length. The effective magnetic field is a sum of several terms: a dc applied magnetic field ($H_0 \hatb{x}$), the Oersted field produced by the electric current in the Pt layer, the demagnetizing field $\vec{H}_\mathrm{dem}(\vec{M})$, the perpendicular anisotropy field, and the exchange field:
\begin{widetext}
\begin{eqnarray}
\vec{H}_\mathrm{eff} & = &  [H_0-H_\mathrm{Oe}^0-\sqrt{2} H_\mathrm{Oe}^\mathrm{ac} \cos (\omega t) ] \hatb{x}+\vec{H}_\mathrm{dem}(\vec{M})  
+\frac{2K_\mathrm{s}}{M_0b} m_s(x) m_y \hatb{y} +\frac{D}{m_s(x)} \frac{\partial}{\partial x} 
(m_s^2(x)  \frac{\partial \vec{m}}{\partial x}) \; ,
\label{Heff}
\end{eqnarray}
\end{widetext}
where $\vec{m}=\vec{M}/M_s(x)$ is the magnetization normalized to its local magnitude $M_s(x)$, $m_s(x) \equiv M_s(x)/M_0$, i.e. with these definitions $\vec{M}=M_0m_s(x)\vec{m}$, $|\vec{m}|=1$. The Oersted field $[-H_{Oe}^0-\sqrt{2} H_{Oe}^{\mathrm{ac}} \cos (\omega t)] \hatb{x}$ is modeled as uniform over the Py wire volume and it is generated by an electric current in Pt: $I_{\text{Pt}}(t)=I_{\text{Pt}}^{\mathrm{dc}}+\sqrt{2} I_{\text{Pt}}^{\mathrm{ac}} \cos (\omega t)$, where $I_{\text{Pt}}^{\mathrm{dc}}$ is direct current in Pt and $I_{\text{Pt}}^{\mathrm{ac}}$ is rms ac current in Pt. Details of the Oersted field model are discussed in the Appendix (section \ref{CO}). The perpendicular anisotropy constant $K_s$ includes contributions from both the top and bottom interfaces of the Py film \cite{Rantschler2005}. $D=2A/M_0$ is the exchange stiffness constant, and $A=5 \times 10^{-7}$\, erg/cm is the exchange constant \cite{Duan2014b}. 

The magnetization dynamics is described by $\vec{m}(x,t)$ through a complex field $a(x,t)$ and its complex conjugate $a(x,t)^*$ via:
\begin{equation}
\begin{array}{ccc}
m_x = 1-aa^* , \\
m_y = -(i/2)(a-a^*)\sqrt{2-aa^*} ,\\ 
m_z = (1/2)(a+a^*)\sqrt{2-aa^*} ,
\label{ma}
\end{array}
\end{equation}
a representation that guarantees $\vec{m}^2(x,t)=1$ everywhere. The Landau-Lifshitz equations of motion, including damping and spin transfer, take a nearly Hamiltonian form in these variables:
\begin{eqnarray}
i\frac{da}{d\tau} & = &  (1-i \alpha) \frac{1}{m_s(x)} \frac{\delta U}{\delta a^*},
\label{EqM1}
\\
i\frac{da^*}{d\tau} & = &  -(1+i \alpha) \frac{1}{m_s(x)} \frac{\delta U^*}{\delta a}
\; .
\label{EqM2}
\end{eqnarray}
These equations are written in scaled variables $U=E/4\pi M_0^2=U_C+iU_{STT}$ and $\tau = 4\pi M_0 |\gamma| t$, where $E=E_C+iE_{STT}$ is the free energy of the system that includes a conservative real part and an imaginary part that describes the action of spin transfer torque.

The conservative part of the free energy $E_C$ consists of a Zeeman term (including the Oersted field), the surface anisotropy term, the exchange term, and the demagnetizing energy terms. The scaled energy terms approximated to quadratic order in the amplitudes $a,a^*$ are given by the following expressions:
\begin{eqnarray}
U_Z & = &  -h_x(\tau) \int dV m_s(x) (1-aa^*) , \label{UZ} \\
U_A & = & -k_s \int dV m_s^2(x) m_y^2 ,  \label{UA}\\
U_X & \simeq & d \int dV m_s^2(x) \nabla a \cdot \nabla a^* , \label{UX}
 \\
U_D & = &  -(1/8\pi M_0^2) \int dV \vec{H}_D(\vec{M}) \cdot \vec{M} , \label{UD}\\
U_{STT} & = & J \int dV m_s^2(x) aa^*. 
\label{USTT}
\end{eqnarray}
In these expressions, $h_x(\tau)=[H_0-H_{Oe}^0-\sqrt{2}H_{Oe}^{\text{ac}}\cos(\Omega \tau)]/4\pi M_0$, $\Omega=\omega/4\pi M_0|\gamma|$,  $k_s=K_s/(4\pi M_0^2 b)$, and $d=l_{ex}^2=D/4\pi M_0=A/2\pi M_0^2$, where the exchange length is $l_{ex}=4.6$\,nm. Expressions for the exchange and dipolar energies expressed via $a$ and $a^*$ are derived in the Appendix (section \ref{sec:appendix}).

We choose the following boundary conditions at the nanowire edges:
\begin{equation}
a|_{x=\pm c}=0 \; .
\label{aBC}
\end{equation}
Also, notice that in our dilution model the magnetization drops to zero at the edges. Then one can show that Eq.\,(\ref{aBC}) leads to:
\begin{equation}
\frac{\partial M_y}{\partial x}|_{x=\pm c}=\frac{\partial M_z}{\partial x}|_{x=\pm c}=0 \; ,
\label{FBC}
\end{equation}
with $M_{x,y}=M_s(x)m_{x,y}$.

A solution of the LLG equations for the complex spin wave amplitude $a(X,\tau)$ that satisfies these boundary conditions can be written as:
\begin{equation}
a(X,\tau) = \sum_{l=1}^N [  a_l (\tau) \cos (k_l X) +f_l(\tau) \sin (q_l X) ],
\label{af}
\end{equation}
where $X \equiv x/c$, $k_l=(2l-1)\pi/2$, and $q_l=l\pi$. 

Linearizing the equations of motion Eqs.\,(\ref{EqM1},\ref{EqM2}) in the absence of ac currents and using the ansatz Eq.\,(\ref{af}), we derive the following equations for the time evolution of the coefficients $a_l(t)$:
\begin{equation}
i \left(
\begin{array}{c}
\dot{a} \\ \dot{a}^*
\end{array}
\right)
= 
\tilde{M}
\left(
\begin{array}{c}
{a} \\ {a}^*
\end{array}
\right) , \label{Eqa}
\end{equation}
where the expression for the matrix $\tilde{M}$ is given by Eq.\,(\ref{Mt}) in the Appendix. 
In Eq.\,(\ref{Eqa}), $a$ is a vector $(a_0,\ldots,a_N)^T$. The equations for $(f_0,\ldots,f_N)^T$ are similar. Notice that due to the symmetry of the system, in the linear approximation the  equations of motion (\ref{EqM1},\ref{EqM2}) separate between even and odd modes, i.e. $\dot{a}_l$ depends only on $a_j$'s and $a_i^*$'s, and similarly for $\dot{f}_l$, i.e. it depends only on $f_j$'s and $f_i^*$'s. 

We seek solutions of Eq.\,(\ref{Eqa}) in the following form:
\begin{equation}
a_l (\tau) = c_l \exp (-i \Omega \tau -\nu \tau)+d_l \exp (i \Omega \tau -\nu \tau) \; .
\label{alt1}
\end{equation}
Substitution of the ansatz Eq.\,(\ref{alt1}) into Eq.\,(\ref{Eqa}) leads to the following eigenvalue problem:
\begin{eqnarray}
\tilde{M}\cdot v = \tilde{\delta} v \; ,
\label{MatEq}
\end{eqnarray}
where $\tilde{\delta} = \Omega -i \nu$ and $v^T=(c^T,(d^*)^T)$. The eigenmodes of this problem, including damping and spin transfer torque, are the right eigenvectors of $\tilde{M}$. A matrix $W$ is constructed with these eigenvectors as its columns, and defines a change of variables to the amplitudes $b_n, b_n^*$ of the eigenmodes as follows:
\begin{eqnarray}
\left( 
\begin{array}{c}
a \\ a^*
\end{array}
\right)
 = W \cdot \left( 
\begin{array}{c}
b \\ b^*
\end{array}
\right) .
\end{eqnarray}
Thus, we obtain the following diagonal equations of motion for the amplitudes of each eigenmode:
\begin{equation}
i \left(
\begin{array}{c}
\dot{b} \\ \dot{b}^*
\end{array}
\right)
= 
\tilde{D} \cdot
\left(
\begin{array}{c}
b \\ b^*
\end{array}
\right) ,
\label{emds}
\end{equation}
with $\tilde{D}=W^{-1} \tilde{M} W$ being a diagonal matrix, whose elements are the frequencies of the modes with associated imaginary parts as decay/growth rates, i.e. $\tilde{\delta_n}=\Omega_n-i\nu_n$. At a critical value of the direct current $I_\mathrm{c}$, the imaginary part of an eigenvalue may go to zero signaling transition of the mode into the regime of auto-oscillations.

For a non-zero ac current generating ac Oersted field and ac spin transfer torque, the equations of motion (\ref{emds}) are modified into: 

\begin{equation}
 i \left(
\begin{array}{c}
\dot{b} \\ \dot{b}^*
\end{array}
\right)
 =
\tilde{D} \cdot
\left(
\begin{array}{c}
b \\ b^*
\end{array}
\right) + \tilde{N}_\mathrm{ac}(\tau) \cdot
\left(
\begin{array}{c}
b \\ b^*
\end{array}
\right) ,
\label{emgo}
\end{equation}
where
\begin{eqnarray}
\tilde{N}_\mathrm{ac}(\tau) & = & 
 W^{-1} \left(
\begin{array}{cc}
(1-i \alpha)H_\mathrm{ac}(\tau) & 0  \\
0 & -(1+i \alpha)H_\mathrm{ac}^*(\tau)
\end{array}
\right)W ,  \nonumber \\ \label{Nac} \\
H_\mathrm{ac}(\tau) &= & -h_\mathrm{ac}(\tau)\mathbb{I}+iJ_\mathrm{ac}(\tau) \tilde{A} , \\
h_\mathrm{ac}(\tau)  & = & h_\mathrm{ac} (e^{i 2\Omega_p \tau}+e^{-i 2\Omega_p \tau})/\sqrt{2} , \\
J_\mathrm{ac}(\tau)  & = & J_\mathrm{ac} (e^{i 2\Omega_p \tau}+e^{-i 2\Omega_p \tau})/\sqrt{2} ,
\end{eqnarray}
where $h_\mathrm{ac}(\tau)$ is the ac Oersted field normalized by $4\pi M_s$, $J_\mathrm{ac}(\tau)$ the ac component of the spin transfer coefficient $J$, which is proportional to the current, $\mathbb{I}$ is a unitary matrix, $\tilde{A}$ a matrix given by Eq.\,(\ref{aln}) of the Appendix. The frequency of the ac current is written as $\Omega=2\Omega_p$ with application of these equations to the analysis of parametric spin wave excitation in mind. 

\subsection{Eigenmodes}

\begin{figure}[ht]
\includegraphics[width=0.99\columnwidth]{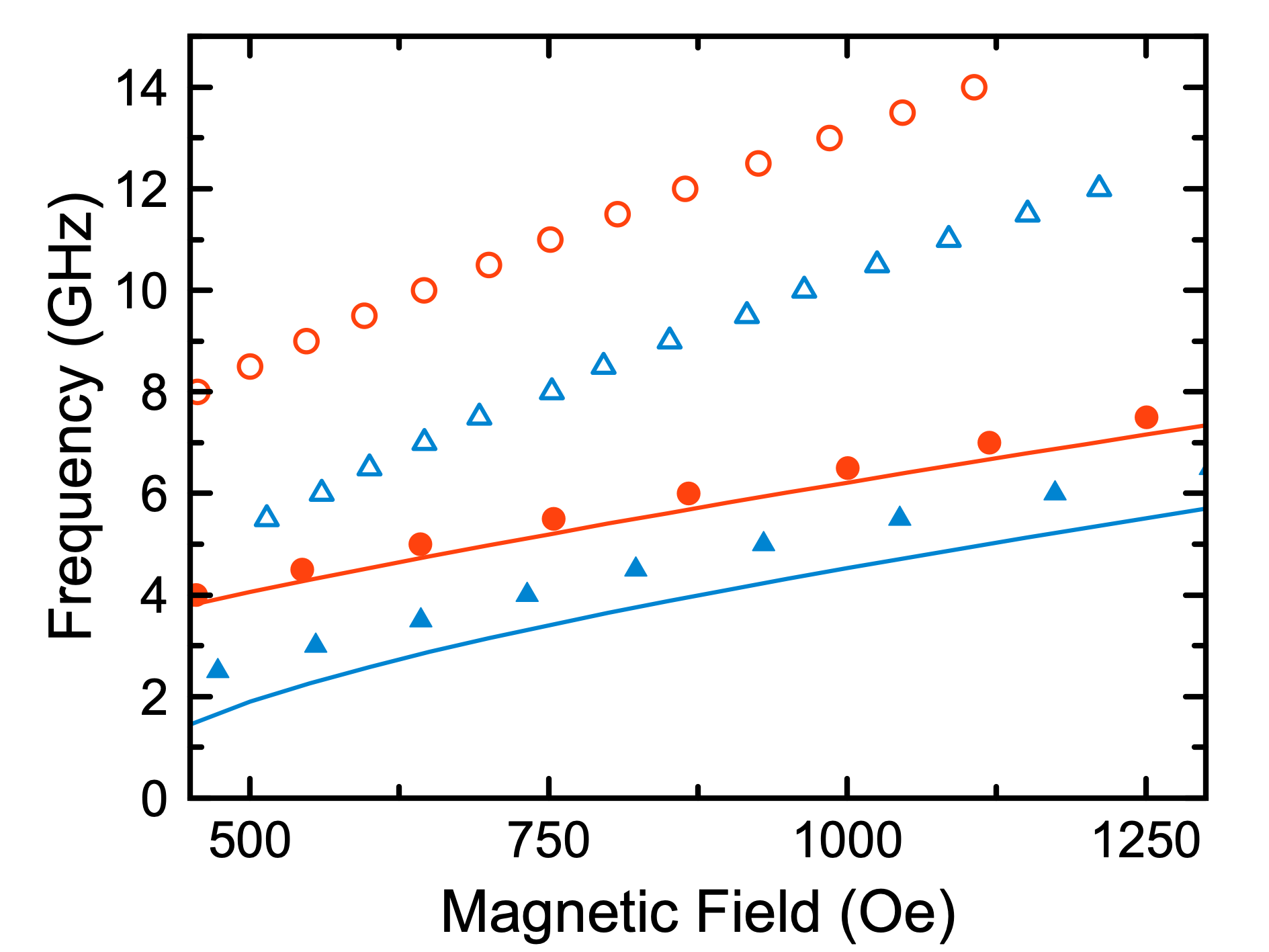}
\caption{Measured (solid symbols) and theoretically calculated (lines) frequencies of the lowest-energy edge (blue) and bulk (red) eigenmodes. Open symbols show measured drive frequencies for parametric excitation of the bulk and edge eigenmodes.}
\label{Fig:borde}
\end{figure}

The spin wave eigenmodes and corresponding eigenfrequencies of the Py nanowire are solutions of Eq.\,(\ref{MatEq}) with the dissipation and spin transfer torque terms set to zero. Lines in Fig.\,\ref{Fig:borde} show the lowest-energy edge and bulk mode frequencies given by Eq.\,(\ref{MatEq}) versus magnetic field  applied in the sample plane perpendicular to the nanowire axis. Note that the edge dilution model is included in our theory. Solid symbols in Fig.\,\ref{Fig:borde} show the dependence of the lowest-energy bulk and edge modes measured by ST-FMR technique. Opens symbols in Fig.\,\ref{Fig:borde} show experimentally measured parametric resonance frequencies of the lowest-energy edge and bulk modes.

Figure~\ref{Fig:borde} reveals good agreement between the measured and calculated values of eigenfrequencies for the lowest-frequency bulk mode. The agreement for the edge mode is substantially worse, indicating that the edge dilution model does not fully capture magnetic properties of the nanowire at the edges. Indeed, since the amplitude of the edge mode is maximized near the wire edge, its frequency is much more sensitive to the magnetic edge properties than the bulk mode. Figure~\ref{Fig:borde} shows that improvements to the edge dilution model used are needed for quantitative description of spin wave eigenmodes in thin-film nanomagnetic elements. We also note that calculations without any edge dilution show much worse agreement with the experiment for the edge eigenmodes, and to a much lesser extent for the bulk eigenmodes. 

Now we turn attention to the spatial profiles of the lowest-energy modes. In the linear approximation, the $M_y, M_z$ components of the modes are given by:
\begin{widetext}
\begin{eqnarray}
    M_y(X,\tau)& = & M_s(X)\sqrt{2}Im(a) 
    = M_s(X)[
    C_I(X)\cos (\Omega \tau)-S_R(X)\sin (\Omega \tau) ] ,
    \label{My}
    \\
    M_z(X,\tau)& = & M_s(X)\sqrt{2}Re(a) 
    = M_s(X)[
    C_R(X)\cos (\Omega \tau)+S_I(X)\sin (\Omega \tau) ] ,
    \label{Mz}
\end{eqnarray}
\end{widetext}
where $C_{R,I}(X)$ and $S_{R,I}(X)$ represent the real (R) and imaginary (I) parts of:
\begin{eqnarray}
 C(X) & = & \sqrt{2} \sum_{l=0}^N (c_l+d_l) \cos ((2l-1)\pi X/2) , \\
 S(X) & = & \sqrt{2} \sum_{l=0}^N (c_l-d_l) \cos ((2l-1)\pi X/2) .
\end{eqnarray}

\begin{figure*} [ht]
\includegraphics[width=0.99\columnwidth]{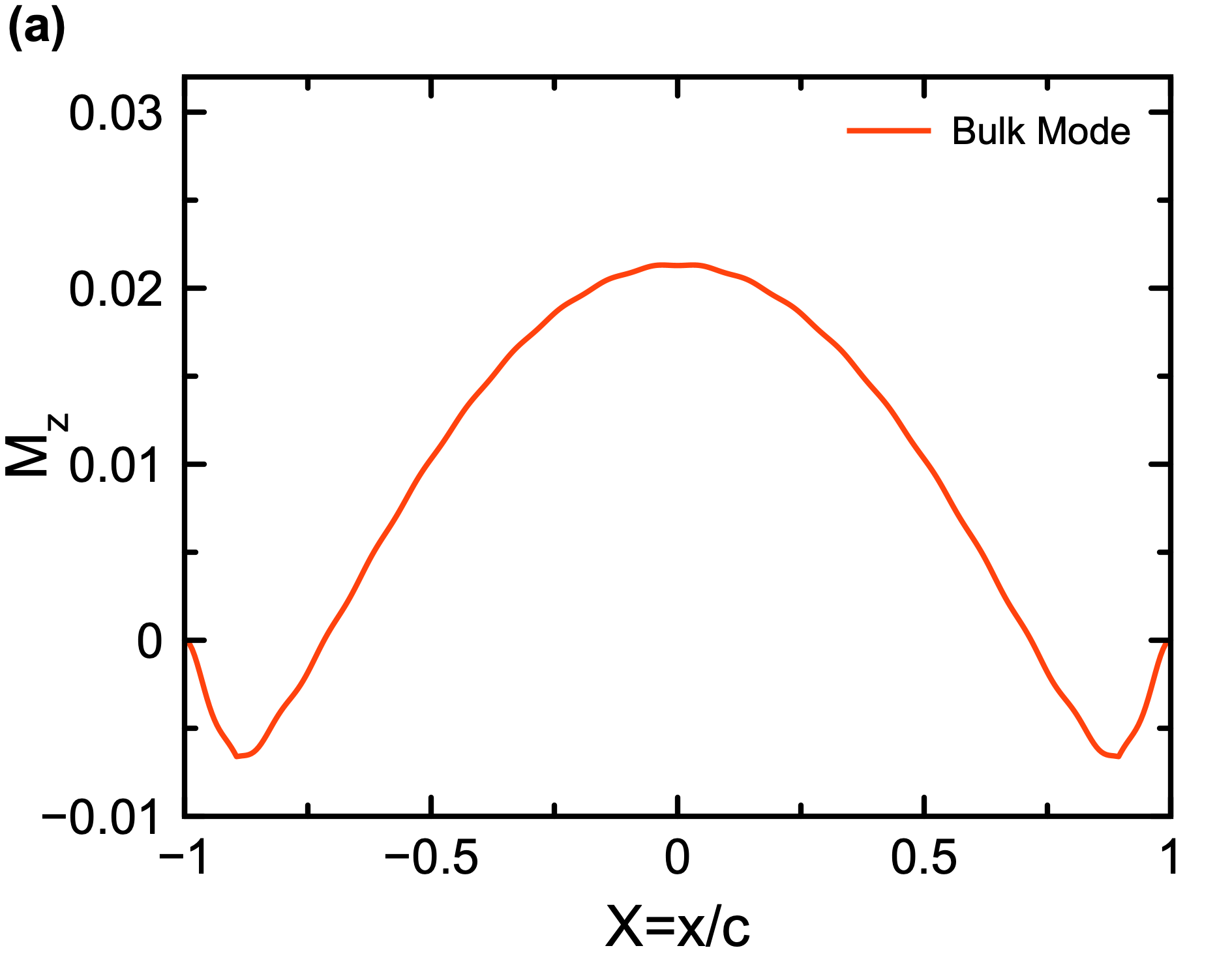}
\includegraphics[width=0.99\columnwidth]{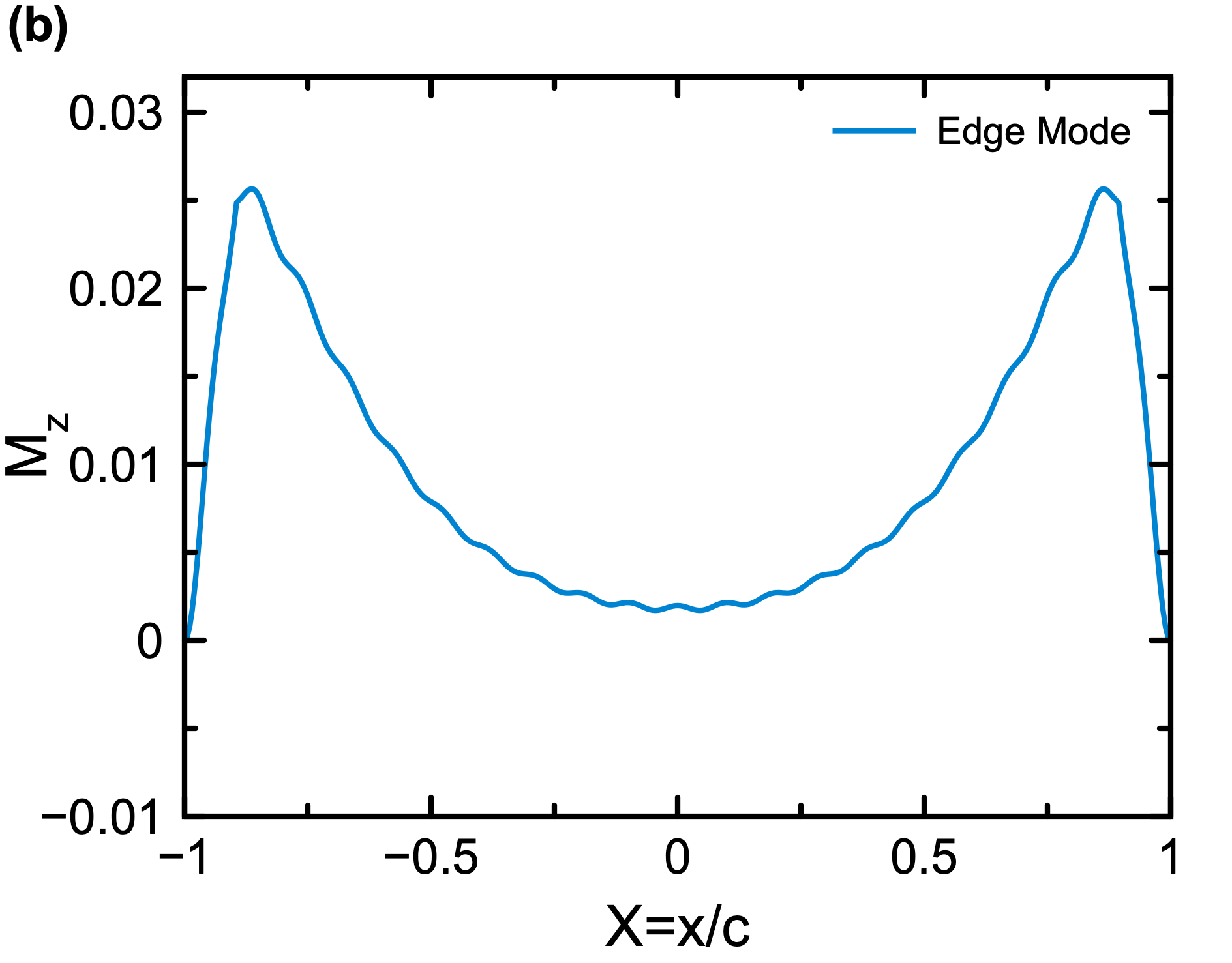}
\caption{Spatial profiles of the dynamic magnetization component amplitude $M_z(X)$ for the lowest-energy bulk (a) and edge (b) modes calculated at $H_0=642.5$\,Oe.}
\label{Fig:shape}
\end{figure*}

Figure~\ref{Fig:shape} shows spatial profiles of the lowest-energy bulk and edge modes at an applied magnetic field $H_0=642.5$\,Oe. We find $C_I(X)=0=S_I(X)$ for both types of modes, which means that $M_y(X,\tau)=-M_s(X)S_R(X)\sin (\Omega \tau)$ and $M_z(X,\tau)=M_s(X)C_R(X)\cos (\Omega \tau)$, i.e. they represent counter-clockwise elliptic precession for both bulk and edge modes. Figures \ref{Fig:shape}(a) and \ref{Fig:shape}(b) show the spatial profiles of the $z$ component of the magnetization (i.e. $M_s(X)C_R(X)$) of the bulk and edge modes respectively. As expected, the bulk mode shows amplitude maximum in the center of the nanowire ($X=0$) while the edge mode has minimum amplitude at $X=0$. The peak amplitude of the edge mode is not located exactly at the wire edge due to the dilution. Furthermore, the ellipticity of these oscillations is defined as \cite{Gurevich1996} $\epsilon = 1-|m_{min}|^2/|m_{max}|^2$ ($|m_{min,max}|$ corresponding to minimum and maximum values at the elliptical axis), which in our case becomes $\epsilon = 1-|S_R(X)|^2/|C_R(X)|^2$. The ellipticity is  approximately 0.75 close to the edges of the stripe and 0.84 in the central part for the bulk mode, while these values are approximately 0.88 and 0.98 respectively for the edge mode, thus the edge mode theoretically shows higher ellipticity than the bulk mode (these values correspond to the modes of Fig.\,\ref{Fig:shape}).  

\subsection{Parametric resonance}

Here we present a simple model describing parametric excitation of spin wave eigenmodes in our nanowire samples by a microwave current at approximately twice the mode frequency. In our model, the frequency of the microwave current is written as $\Omega = 2 \Omega_p$,  where $\Omega_p$ is similar to the eigenmode frequency $\Omega_n$. In the equations of motion Eq.\,(\ref{emgo}), we focus on a single mode of index $n$ and neglect all non-resonant terms:
\begin{equation}
i \dot{b}_n = \tilde{\delta}_n b_n + N_{n n^*} b_n^* e^{-i 2 \Omega_p \tau}/\sqrt{2} \; ,
\label{eappo}
\end{equation}
where $\tilde{\delta_n}=\Omega_n-i\nu_n$. A similar equation is written for $b_n^*$.
We seek a solution of these equations in the following form:
\begin{equation}
\begin{array}{ccc}
b_n \simeq  b_n^0 e^{-i \Omega_p \tau -\nu \tau} .
\end{array}
\label{blao}
\end{equation}
Inserting Eq.\,(\ref{blao}) and its complex conjugate into Eq.\,(\ref{eappo}) and a similar equation for for $b_n^*$ leads to the following set of homogeneous linear algebraic equations:
\begin{widetext}
\begin{eqnarray}
& & \left(
\begin{array}{cc}
(\Omega_p-\Omega_n)-i(\nu-\nu_n) & 
-N_{n n^*}/\sqrt{2} \\
-(N_{n n^*})^*/\sqrt{2}    & (\Omega_p-\Omega_n)+i(\nu-\nu_n)
\end{array}
\right)
\left(
\begin{array}{c}
b_n^0 \\
b_n^{0*}
\end{array}
\right) = 0 .
\label{detcon}
\end{eqnarray}
\end{widetext}
The non-trivial solution of Eq.\,(\ref{detcon}) is found from the zero determinant condition, i.e.:
\begin{equation}
|\nu-\nu_n |= \sqrt{|N_{n n^*}|^2/2-(\Omega_p-\Omega_n)^2}.
\end{equation}
A steady state oscillatory solution of Eq.\,(\ref{detcon}), i.e. $\nu=0$, is given by the following condition for $|N_{n n^*}|$:
\begin{equation}
|\nu_n| = \sqrt{|N_{n n^*}|^2/2-(\Omega_p-\Omega_n)^2} . \label{cond}
\end{equation}
Since $N_{n n^*}$ is proportional to the ac current $I_\mathrm{ac}$, we can write it as $ N_{n n^*} = I_\mathrm{ac} \hat{N}_{n n^*} $, where $\hat{N}_{n n^*}$ is a current-independent coefficient. It is clear that the minimum ac current that satisfies Eq.\,(\ref{cond}) is achieved for $\Omega_p=\Omega_n$, when $|\nu_n| = |N_{n n^*}|/\sqrt{2}$. This gives us an expression for the threshold ac current for excitation of parametric resonance for a given mode:
\begin{equation}
I_\mathrm{th} = \sqrt{2}|\nu_n|/|\hat{N}_{nn^*}| .
\label{ccn}
\end{equation}

The matrix element $\hat{N}_{nn^*}$ can be obtained from Eq.\,(\ref{Nac}). The ac current is given by $I_\mathrm{ac}(\tau) = I_\mathrm{ac} (e^{i 2\Omega_p \tau}+e^{-i 2\Omega_p \tau})/\sqrt{2}$, and Eq.\,(\ref{Nac}) can be rewritten to explicitly factor out $I_\mathrm{ac}(\tau)$:
\begin{widetext}
\begin{eqnarray}
& \tilde{N}_\mathrm{ac}(\tau) = I_\mathrm{ac} (\tau) \hat{N} = I_\mathrm{ac} (\tau)
\left[ W^{-1} Y W \right]
= I_\mathrm{ac} (\tau)
\left[ W^{-1} \left(
\begin{array}{cc}
(1-i \alpha)(-k \mathbb{I}+i \beta \tilde{A}) & 0  \\
0 & (1+i \alpha)(k\mathbb{I}+i\beta \tilde{A})
\end{array}
\right)W \right] ,
\nonumber \\ \label{Nh}
\end{eqnarray}
\end{widetext}
where we have used the fact that both the ac Oersted field $h_\mathrm{ac}$ and ac spin transfer torque described by $J_\mathrm{ac}$ are proportional to $I_\mathrm{ac}$, and have written them as $h_\mathrm{ac}=k I_\mathrm{ac}$ and $J_\mathrm{ac}=\beta I_\mathrm{ac}$ ($\mathbb{I}$ is the unit matrix). The coefficient $\hat{N}_{nn^*}$ determining the value of $I_\mathrm{th}$ is then the element $(nn^*)$ of the matrix $\hat{N}$ in Eq.\,(\ref{Nh}).

Thus, the expression of Eq.\,(\ref{ccn}) for the threshold rms ac current for parametric excitation $I_\mathrm{th}$  depends on two quantities,  $|\nu_n|$ and $|\hat{N}_{nn^*}|$, that exhibit different dependence on $I_\mathrm{dc}$: $|\nu_n|$ dependence on $I_\mathrm{dc}$ is approximately linear, while $|\hat{N}_{nn^*}|$ dependence on $I_\mathrm{dc}$ is weak. This explains the linear dependence of $I_\mathrm{th}$ on $I_\mathrm{dc}$ observed experimentally in Fig.\,\ref{Fig:FittedRFthreshold}. 

Indeed, using Eq.\,(\ref{Mt}) from the Appendix for $\tilde{M}$ of Eq.\,(\ref{Eqa}), we can write:
\begin{equation}
    \tilde{M} = M +i J \left(
    \begin{array}{cc}
    \tilde{A} & 0 \\ 0 & \tilde{A}
    \end{array}
    \right)
    \simeq  M +i J 1 \; ,
    \label{Mts}
\end{equation}
where $M$ does not depend on spin torque, and $1$ is a unit matrix (of a double size compared to $\mathbb{I}$). The last approximation in Eq.\,(\ref{Mts}), that assumes a diagonal form of $\tilde{A}$ and that is valid for zero edge dilution, is a better approximation for the bulk modes than for the edge modes. Within the latter approximation, an eigenvector of the matrix $M$ with eigenvalue $\delta_n=\Omega_n-i\nu_n^0$ is an eigenvector of the matrix $\tilde{M}$ with eigenvalue $\tilde{\delta}_n=\Omega_n-i(\nu_n^0-J)$. Here $\nu_n^0$ is the decay constant of mode $n$ at zero spin transfer torque; $\nu_n^0$ is approximately independent of current (it depends slightly on dc current through the effective applied magnetic field modified by the Oersted contribution). The approximate expression $\tilde{\delta}_n=\Omega_n-i(\nu_n^0-J)$ validates linear behavior of $\nu_n \simeq \nu_n^0-J$ on $I_\mathrm{dc} \sim J$. Furthermore, from Eq.\,(\ref{Nh}) we can show that $|\hat{N}_{nn^*}|$ is approximately independent of $I_\mathrm{dc}$: the matrix $Y$ depends on parameters independent of $I_\mathrm{dc}$ and the eigenvectors that form the matrix W only weakly depend on $I_\mathrm{dc}$. 

We note that $Y$ in Eq.\,(\ref{Nh}) depends on a linear combination of the parameters defining the efficiencies of the Oersted field ($k$) and antidamping spin torque ($\beta$):  $(\mp k\mathbb{I}+i\beta \tilde{A})$. This means that both the Oersted field and spin torque contribute to the excitation of parametric resonance on qualitatively equal footing. However, our theoretical analysis below reveals that for the materials and geometry considered in this paper, the contribution of the Oersted field to the excitation of parametric resonance is dominant over that of spin Hall torque. For example, if we artificially turn off the ac Oersted field ($k=0$) for the bulk mode at $I_\mathrm{dc}=0$, we calculate $|\hat{N}_{nn*}|=0.0002$, while if we artificially turn off the ac spin transfer term ($\beta=0$), $|\hat{N}_{nn*}|=0.0040$. This implies via Eq.\,(\ref{ccn}) that the ac Oersted field comprises approximately 95\% of the parametric resonance drive.

A qualitative explanation of the dominant role of the Oersted field is given in Appendix \ref{tfl} where we derive analytical expressions for a simple case of the uniform mode of precession in the limit of infinite Py/Pt bilayer. This example allows us to qualitatively understand why the ac Oersted field is the dominant parametric drive for the more general case of the Py/Pt bilayer nanowire. The matrix $Y$ in Eq.\,(\ref{Nh}) can be separated into a term proportional $k$ and a term proportional to $\beta$. In this case the dominant contribution to the term proportional to $\beta$ can be estimated by taking $\alpha=0$ (the low damping limit $\alpha\ll 1$) and $\tilde{A}= \mathbb{I}$ (zero edge dilution, as there are no edges for the infinite bilayer). Under these approximations, the term proportional to $\beta$ becomes $i\beta W^{-1}W=i\beta 1$. When this purely imaginary and uniform diagonal matrix is used in the equation of motion Eq.\,(\ref{emgo}), it does not generate any coupling between $b$ and $b^*$, which implies infinite threshold for parametric excitation under the purely spin torque drive in this approximation. Another consequence of the theoretical model that points in the direction of explaining the preponderance of the Oersted field in parametric resonance in this experiment is that without dilution the spin transfer torque term does not produce a coupling between $b_n$ and $b_n^*$ for all modes. The latter happens because the expression for the imaginary energy associated to spin transfer of Eq.\,(\ref{USTT}) is proportional to $\sum_l a_l a_l^*$, and this "diagonal" property persists in terms of the variables $b_n, b_n^*$, meaning that spin transfer does not couple $b_n$ with $b_n^*$, i.e. it does not contribute to parametric resonance excitation.

\section{\label{sec:discussion}Discussion} 

In this section we compare experimental results to theoretical predictions, focusing on the dependence of $I_\mathrm{th}$ on $I_\mathrm{dc}$ in Fig.\,\ref{Fig:FittedRFthreshold}(c). In particular, we compare the experimental slopes of the $I_\mathrm{th}(I_\mathrm{dc})$ linear dependence in Fig.\,\ref{Fig:FittedRFthreshold}(c) for the lowest energy bulk and edge modes with the theoretical slopes for those modes. Since both $I_\mathrm{th}$ and $I_\mathrm{c}$ are linear in the mode damping constant, the slope of $I_\mathrm{th}(I_\mathrm{dc})$ is independent on the damping and primarily characterizes ellipticity of the mode. Indeed, at a fixed mode frequency, $I_\mathrm{th}$ decreases with increasing mode ellipticity \cite{Chen2017} while $I_\mathrm{c}$ increases with increasing mode ellipticity \cite{Grollier2003,Chen2011}, which makes the slope of $I_\mathrm{th}(I_\mathrm{dc})$  a very sensitive probe of the mode ellipticity. We use this probe to test the theoretical description of the bulk and edge spin wave eigenmodes.

The theoretical results of Eqs.\,(\ref{ccn}) and (\ref{Nh}) allow us to calculate $I_\mathrm{th}$ at a given $I_\mathrm{dc}$. The theoretical slopes are found by calculating the intercept points of the $I_\mathrm{th}(I_\mathrm{dc})$ dependence with the abscissa and ordinate: $I_\mathrm{th}(0)\equiv I_\mathrm{th}^0$ and $I_\mathrm{th}(I_\mathrm{c})=0$, where $I_\mathrm{c}$ is the critical current for the onset of auto-oscillations. The absolute value of the slope of $I_\mathrm{th}(I_\mathrm{dc})$ is then given by $I_\mathrm{th}^0/I_\mathrm{c}$.

In order to calculate $I_\mathrm{th}^0$ from Eq.\,(\ref{Nh}), we need first to determine the values of $k$ and $\beta$, which are proportional to the strengths of the Oersted field and spin transfer torque respectively. The constant $k=0.00324$\,mA$^{-1}$ is calculated in the Appendix \ref{CO}. The constant $\beta$ proportional to the spin Hall torque efficiency is determined from the experimentally measured value of the critical current $I_\mathrm{c}$ for the onset of the mode auto-oscillations driven by $I_\mathrm{dc}$ at $I_\mathrm{ac}=0$. This is done by solving Eq.\,(\ref{MatEq}) for $\beta$ with $\nu=0$ and values of the damping constant $\alpha$ and $I_\mathrm{c}$ appropriate for the given mode. In this solution, we use the applied magnetic field value $H_0$ appropriate for the mode frequency of 10\,GHz used in Fig.\,\ref{Fig:FittedRFthreshold}(c). The measured values of $H_0$ at 10\,GHz are given by Fig.\,\ref{Fig:VoltageSpectra}: $H_0=642.5$\,Oe for the bulk mode and $H_0=930$\,Oe for the edge mode.

We solve Eqs.\,(\ref{ccn}) and (\ref{Nh}) to fit the theoretical slope $I_\mathrm{th}^0/I_\mathrm{c}$ to its experimentally measured value $I_\mathrm{th}^0/I_\mathrm{c}=1.88$ in Fig.\,\ref{Fig:FittedRFthreshold}(c) with the damping $\alpha$ as the single fitting parameter. In this fitting procedure, $\nu_n(\alpha)$ in Eq.\,(\ref{ccn}) is found via diagonalization of the matrix $\tilde{M}$ given by Eq.\,(\ref{Mt}) in the Appendix, and $\beta(\alpha)$ is found by solving Eq.\,(\ref{MatEq}) with $\nu=0$ at  $I_\mathrm{dc}=I_\mathrm{c}=2.41$\,mA and $I_\mathrm{ac}=0$. The best fit of the Eq.\,(\ref{ccn}) to the experimental value of $I_\mathrm{th}^0/I_\mathrm{c}=1.88$ for the lowest energy bulk mode in Fig.\,\ref{Fig:FittedRFthreshold}(c) gives $\alpha=0.034$. This value of $\alpha$ is very close to the value of $\alpha=0.031$ directly measured by ST-FMR using the data in Fig.\,\ref{Fig:FMR}. This validates our theoretical model of  the bulk spin wave modes and their excitation by the parametric drive. The fitting procedure gives $|\nu_n|=0.012$ at $I_\mathrm{dc}=0$,  $\beta=0.0047$\,mA$^{-1}$, and $|\hat{N}_{nn^*}|=0.0038$ at $I_\mathrm{dc}=0$ for the lowest energy bulk mode. Notice that this value of $\beta$ allows us to calculate the spin Hall angle as $\theta_H=0.042$, which is consistent with previously reported values in similar devices \cite{Sagasta2016} (see Appendix \ref{sha}).

Using the same value of the damping parameter as for the bulk mode, $\alpha = 0.034$, we calculate for the lowest energy edge mode: $\nu_n(\alpha)=0.010$, $\beta=0.0041$\,mA$^{-1}$ from imposing $\nu=0$ at $I_c=2.59$\,mA in Eq.\,(\ref{MatEq}), and $|\hat{N}_{nn^*}|=0.0035$ at $I_\mathrm{dc}=0$, which gives theoretical value of $I_\mathrm{th}^0=4.04$\,mA. Thus, the theoretically expected slope for the edge mode $I_\mathrm{th}^0/I_c=1.56$ is much lower than its experimentally measured value $2.29$ in Fig.\,\ref{Fig:FittedRFthreshold}(c). Since the slope is not sensitive to the damping constant, the discrepancy between theory and experiment demonstrates that ellipticity of the edge mode predicted by the theory is approximately 40\% higher than that inferred from the experimental data.

Our experimental observation of the lower than expected edge mode ellipticity points to deficiencies of the edge dilution model we use. While the model is a significant improvement over the spatially uniform magnetization model, it does not fully capture the edge magnetization dynamics. We thus conclude that further improvements of the edge dilution model are needed to adequately describe magnetization at the edges of thin-film nanomagnetic structures. We note that this problem is of significant technological relevance because spin transfer torque memory (STT-MRAM) cells are projected to scale down to lateral dimensions below 10\,nm in the near future \cite{Bhatti2017}, which implies that its switching properties will be dominated by the state and dynamics of magnetization at the element edges.

It is important to understand whether the discrepancy between theory and experiment is a result of mathematical approximations employed in the model or has its roots in the physical properties of the magnetic material at the magnetic film edge. For example, can the observed discrepancy be a result of the boundary conditions for dynamic magnetization chosen in the model? In the model, we use the boundary conditions given by Eq.\,(\ref{aBC}) so that the dynamic field $a(x,t)$ is zero at the edges, which leads to free boundary conditions for $M_y$ and $M_z$ [Eq.\,(\ref{FBC})]. To understand the impact of these boundary conditions, we repeated the calculations assuming that the dynamic field $a(x,t)$ has zero derivative at the edges. These calculations show negligible impact on $I_\mathrm{th}^0/I_c$ for the bulk mode, and the change in $I_\mathrm{th}^0/I_c$ for the edge mode is much too small to explain the discrepancy between theory and experiment. The smallness of the impact of the boundary conditions for $a(x,t)$ on the simulation results is reasonable because the edge dilution model used imposes the magnetization to be zero exactly at the edge, and thus boundary condition for the field $a(x,t)$ have little impact on the magnetization dynamics. 

The unexpectedly low ellipticity of the edge mode seen in the experiment is likely to have a physical origin. For example, it can be explained by magnetic anisotropy at the wire edges. Two types of edge magnetic anisotropy can result in decreased ellipticity of the edge mode. First, the perpendicular magnetic anisotropy $K_\mathrm{s}$ at the edge can be enhanced due to Py and Pt intermixing induced by ion milling in the nanowire fabrication process or by partial Py oxidation at the edges \cite{McMichael2010}. This type of anisotropy would indeed decrease the edge mode ellipticity but it would also decrease the mode frequency, bringing it farther away from that seen in the experiment.

Alternatively, a surface magnetic anisotropy with an easy axis perpendicular to the nanowire edge (along the $x$-axis in Fig.\,\ref{Fig:Layout}(a)) \cite{Rantschler2005} can reduce the edge mode ellipticity. Such anisotropy can both reduce the mode ellipticity and increase the mode frequency in agreement with our experimental data. This type of anisotropy can only be non-zero in a modified edge dilution model where magnetization is not reduced to zero at the wire edge.

Another possible explanation of the observed reduced edge mode ellipticity is nanowire edge roughness. It has been previously shown \cite{Cowburn2000} that edge roughness significantly reduces the edge saturation field due to dipolar interactions via the so-called lateral magnetic anisotropy \cite{Cowburn2000}, and thus edge roughness is expected to increase the edge mode frequency. Dipolar interactions arising from edge roughness are also expected to decrease the edge mode ellipticity and thus the edge roughness model can potentially explain all our data. Therefore, development of a mathematical model of edge mode dynamics in the presence of edge roughness is a promising future direction of research. 

We believe that definitive understanding of magnetic properties at the edge of magnetic thin-film elements requires direct imaging of structural and magnetic properties of the edge with atomic resolution, which presents a significant technical challenge. Until such full quantitative characterization is achieved, our results on ellipticity of the edge mode via studies of parametric resonance controlled by antidamping spin Hall torque can serve as a test for future improved models of  magnetic edge modification \cite{Shinozaki2020, Herrera2020}. The novelty of our work compared to prior studies of parametric resonance in magnetic nanostructures \cite{Urazhdin2010,Ulrichs2011,Edwards2012,Akerman2014,Chen2017,Heinz2021,Cho2021} is (i) first measurement and quantitative theoretical understanding of parametric resonance of the edge mode and (ii) development of analytical theory of parametric resonance of spin waves in the nanowire geometry.

\section{\label{sec:conclusions}Conclusions} 

In summary, we have demonstrated parametric excitation of bulk and edge spin wave modes in transversely magnetized Pt/Py bilayer nanowires by a microwave current. The threshold current for the parametric excitation is tunable by direct current bias via the antidamping spin Hall torque, and analysis of the threshold current dependence on spin Hall torque allows us to probe ellipticity of the spin wave modes.

We have developed an analytical theory of the spin wave mode spectrum in the nanowire geometry and parametric excitation of these spin waves by microwave current. Our theory takes into account a model describing dilution of magnetization of Py near the wide edges. Comparison between this theory and experiment shows that our theory provides accurate quantitative description of the bulk spin wave mode properties, including their frequency and ellipticity.

In contrast, the theory significantly underestimates the frequency of the edge spin wave modes and overestimates their ellipticity. This suggest that the edge dilution model used here does not completely capture the magnetic properties of the edge and further refinements of the model are needed to achieve a quantitative description of magnetization dynamics at edges of thin magnetic elements. We have identified inclusion of edge roughness effects as a promising direction for future improvements of the model describing magnetization dynamics at edges of thin magnetic elements. Indeed, edge roughness is expected to increase the edge mode frequency and decrease its ellipticity via the lateral magnetic anisotropy \cite{Cowburn2000}, bringing both of these quantities closer to the experimentally observed values. Further quantitative studies are needed to test if lateral magnetic anisotropy completely describes magnetization dynamics at the nanowire edges.

\section{\label{sec:conclusion}Acknowledgements}

This work was supported by the National Science Foundation through Awards No. EFMA-1641989 and No. ECCS-1708885. We also acknowledge support by the Army Research Office through Award No. W911NF-16-1-0472. R.E.A. acknowledges support by Fondecyt Project 1200829 (Chile), and Basal Program for Centers of Excellence, Grant AFB 180001 CEDENNA, ANID (Chile). A.A.J. acknowledges the financial support of ANID FONDECYT Postdoctorado 3190632 (Chile). The authors acknowledge the use of facilities and instrumentation at the UC Irvine Materials Research Institute (IMRI), which is supported in part by the National Science Foundation through the UC Irvine Materials Research Science and Engineering Center (DMR-2011967). The authors also acknowledge the use of facilities and instrumentation at the Integrated Nanosystems Research Facility (INRF) in the Samueli School of Engineering at the University of California Irvine.

\section{\label{sec:appendix}Appendix}

\subsection{ST-FMR signal: \label{STFMR}}

\begin{figure}[ht]
\includegraphics[width=0.99\columnwidth]{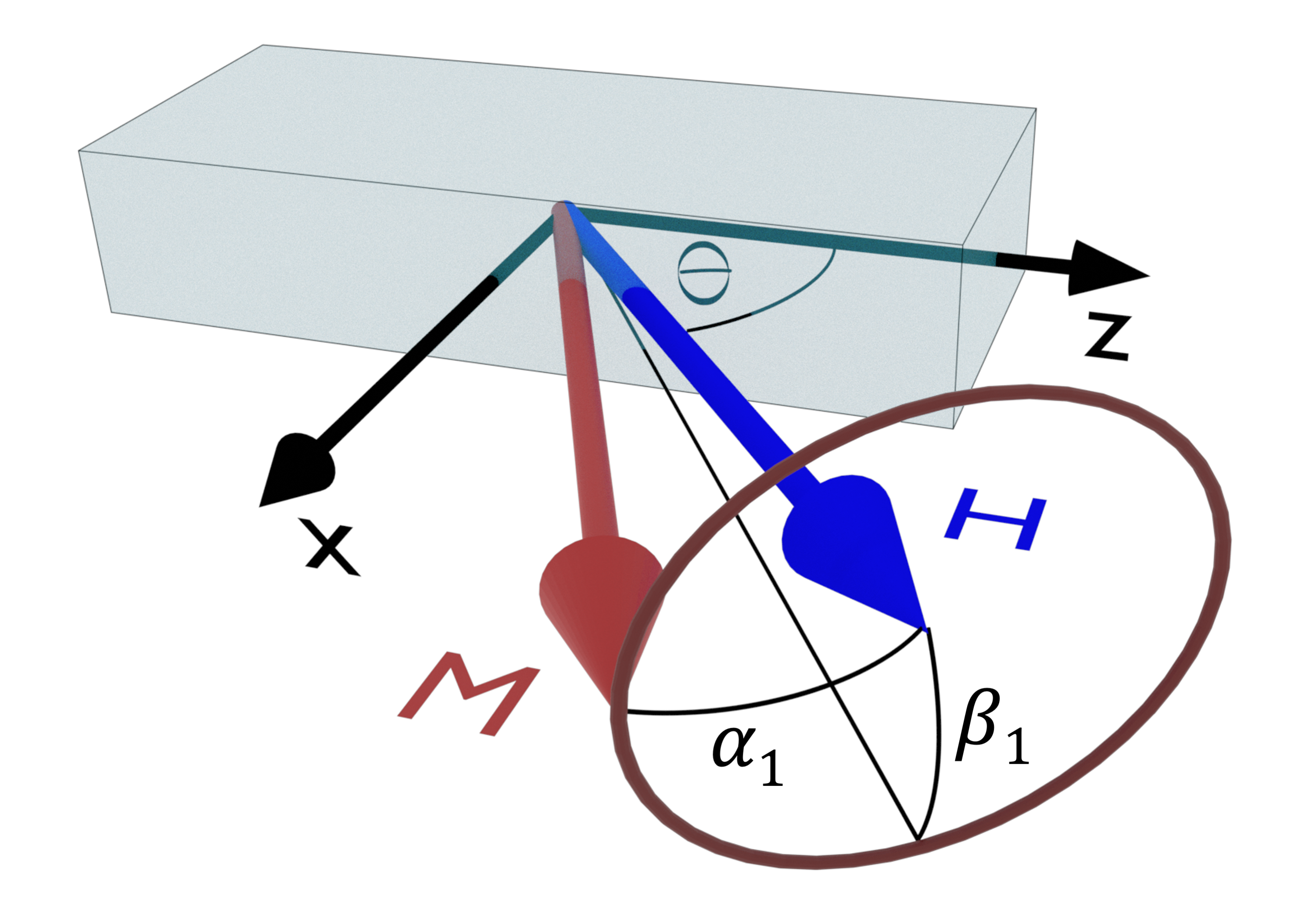}
\caption{Elliptical magnetization precession cone around in-plane magnetic field $H$ applied at angle $\theta$ with respect to the wire axis ($z$-axis). The precession cone is characterized by the major and minor axis cone angles $\alpha_1$ and $\beta_1$.}
\label{Fig:angles}
\end{figure}

The direct voltage $V$ across the sample subjected to a microwave and a direct current consists of three terms: 
\begin{eqnarray}
    V = I_\mathrm{dc}\cdot R_\mathrm{dc}+I_\mathrm{dc}\Delta R_\mathrm{ac}+U_\mathrm{mix}.
\end{eqnarray}
The first term proportional to $I_\mathrm{dc}$ is independent of magnetization dynamics and is simply given by the equilibrium sample resistance $R_\mathrm{dc}$. The second term is the photo-resistance contribution \cite{Mecking2007}, which is proportional to $I_\mathrm{dc}$ and time-averaged change in sample resistance $\Delta R_\mathrm{ac}$ induced by magnetization precession. The third term called photo-voltage \cite{Mecking2007} is the rectified voltage arising from mixing of resistance oscillations and microwave current. The direct voltage $V$ can be calculated as the time-averaged $\langle...\rangle$ total voltage: $V = \langle U(t) \rangle=\langle R(t)\cdot I(t)\rangle$, where $U(t)$, $R(t)$, and $I(t)=I_\mathrm{dc}+\sqrt{2}I_\mathrm{ac}\,\cos(\omega t)$ are the time dependent voltage, resistance, and current. 

Here we derive the direct voltage signal in the configuration of a Py nanowire for both direct excitation and parametric excitation following the approach outlined in Ref.~\cite{Mecking2007}. The time-dependent resistance is given by $R(t)=R_0+R_A\cos^2\phi(t)$, where $R_A$ is the magnitude of AMR, $\phi(t)$ is the instantaneous angle between $\vec{M}$ and the current direction $\hatb{z}$, as shown in Fig.\,\ref{Fig:angles}:
\begin{equation}
\cos\,\phi(t)=\cos\,\alpha(t)\,\cos\,\beta(t),
\label{cos_fi}
\end{equation}
where $\alpha(t)=\theta+\alpha_1\, \cos(\omega t-\psi)$, is the angle between the projection of magnetization onto the $xz$ plane and $z$ axis and $\beta(t)=-\beta_1\, \sin(\omega t-\psi)$ is the tilt angle of magnetization out of the $xz$ plane, as illustrated in Fig.\,\ref{Fig:angles}. In these expressions, $\psi$ is the phase shift between the microwave drive and magnetization oscillations, $\alpha_1$ is the in-plane magnetization oscillation amplitude while $\beta_1$ is the out-of-plane oscillation amplitude. Using these expressions in Eq.\,(\ref{cos_fi}), we expand $\cos^2\phi(t)$ to second order in $\alpha_1$ and $\beta_1$ \cite{Mecking2007}:
\begin{eqnarray}
\cos^2\,\phi(t) &=& \cos^2\,\theta - \alpha_1\, \sin\, 2\theta\, \cos(\omega t-\psi) \nonumber \\
&-&\alpha^2_1\, \cos\, 2\theta\, \cos^2(\omega t-\psi)  \nonumber \\
&-& \beta^2_1\, \cos^2\theta\, \sin^2(\omega t - \psi).
\label{cos2fi}
\end{eqnarray}

For direct (linear) excitation of a spin wave eigenmode by a microwave current at the eigenmode frequency $\omega$, the time-dependent voltage across the sample $U_\mathrm{lin}(t)$ is:
\begin{eqnarray}
U_\mathrm{lin}(t) &=& [R_0+R_A\,\cos^2\phi(t)][I_\mathrm{dc}+\sqrt{2}I_\mathrm{ac}\,\cos(\omega t)]. \nonumber \\
\end{eqnarray}

Using Eq.\,(\ref{cos2fi}) in this latter expression and calculating the time average of $U_\mathrm{lin}(t)$, we obtain the direct voltage $V_\mathrm{lin}$ across the sample:
\begin{eqnarray}
V_\mathrm{lin} &=& I_\mathrm{dc}(R_0+R_A\,\cos^2\theta)\\ \nonumber
&& -\frac{1}{2}I_\mathrm{dc}R_A(\alpha^2_1\cos\,2\theta+\beta^2_1\cos^2\theta)\\ \nonumber
&& -\frac{\sqrt{2}}{2}I_\mathrm{ac}R_A\alpha_1\,\sin\, 2\theta\,\cos(\psi),
\end{eqnarray}
where the first term  $I_\mathrm{dc}(R_0+R_A\,\cos^2\theta)$ is the equilibrium direct voltage independent of spin wave excitation, the second term $-\frac{1}{2}I_\mathrm{dc}R_A(\alpha^2_1\cos\,2\theta+\beta^2_1\,\cos^2\theta)$ is the photo-resistance term proportional to $I_\mathrm{dc}$, and the last term  $-\frac{\sqrt{2}}{2}I_\mathrm{ac}R_A\alpha_1\sin\,2\theta\cos(\psi)$ is the photo-voltage term proportional to $I_\mathrm{ac}$. 

For parametric excitation of a spin wave eigenmode, we use a microwave current at twice the eigenmode frequency: $I(t)=I_\mathrm{dc}+\sqrt{2}I_\mathrm{ac}\,\cos(2\omega t)$. Therefore, the time-dependent voltage across the sample $U_\mathrm{par}(t)$ is
\begin{eqnarray}
U_\mathrm{par}(t) &=& [R_0+R_A\,\cos^2\phi(t)][I_\mathrm{dc}+\sqrt{2}I_\mathrm{ac}\,\cos(2\omega t)].\nonumber
\end{eqnarray}

Using Eq.\,(\ref{cos2fi}) in this expression and calculating the time average of $U_\mathrm{par}(t)$, we obtain the direct voltage $V_\mathrm{par}$ across the sample:
\begin{eqnarray}
V_\mathrm{par}&=& I_\mathrm{dc}(R_0+R_A\,\cos^2\theta)\\ \nonumber
&& -\frac{1}{2}I_\mathrm{dc}R_A(\alpha^2_1\cos\,2\theta+\beta^2_1\cos^2\theta)\\ \nonumber
&& -\frac{\sqrt{2}}{4}I_\mathrm{ac}R_A(\alpha^2_1\cos\,2\theta+\beta^2_1\cos^2\theta)\cos(2\psi).
\end{eqnarray}

In the experimental configuration used in this work $\theta = \pi/2$, and thus $V_\mathrm{lin}$ and $V_\mathrm{par}$ can be further simplified:
\begin{eqnarray}
V_\mathrm{lin} &=& I_\mathrm{dc}R_0 +\frac{1}{2}I_\mathrm{dc}R_A\alpha^2_1,
\label{eq:direct_90deg}
\end{eqnarray}
\begin{eqnarray}
V_\mathrm{par} &=& I_\mathrm{dc}R_0 +\frac{1}{2}I_\mathrm{dc}R_A\alpha^2_1 +\frac{\sqrt{2}}{4}I_\mathrm{ac}R_A\alpha^2_1\cos(2\psi). \qquad 
\label{eq:parametric_90deg}
\end{eqnarray}

For our device geometry, the phase shift $\psi \approx 0$. Therefore, we can simplify Eq.\,(\ref{eq:parametric_90deg}) by setting $\psi=0$.

We can further use results of Ref.~\cite{Chen2017}, where expressions for current-driven parametric resonance amplitude $\propto \alpha_1$ and power $\propto \alpha_1^2$ were derived in the limits of the microwave drive amplitude ($I_\mathrm{ac}$) well below and well above the threshold drive for parametric excitation ($I_\mathrm{th}$):

\begin{eqnarray}
    \alpha_1^2 = 
    \begin{cases}
      A/(I_\mathrm{ac}-I_\mathrm{th})^2 & \text{$I_\mathrm{ac}\ll I_\mathrm{th}$ }\\
      B\sqrt{I_\mathrm{ac}^2-I_\mathrm{th}^2} & \text{$I_\mathrm{ac}\gg I_\mathrm{th}$, }
    \end{cases}
    \label{parpower}
\end{eqnarray}
where A and B are constants. In Eq.\,(\ref{parpower}), the amplitude of precession below $I_\mathrm{th}$ is not zero due to thermally assisted excitation of the spin wave eigenmode \cite{Chen2017}. 

Using Eq.\,(\ref{parpower}) in Eq.\,(\ref{eq:parametric_90deg}), we calculate the expression for direct voltage arising form parametric excitation of a spin wave eigenmode:
\begin{eqnarray}
      V_\mathrm{par} \sim 
    \begin{cases}
      \left(2I_\mathrm{dc}+\sqrt{2}I_\mathrm{ac}\right)/(I_\mathrm{th}-I_\mathrm{ac})^2 & \text{$I_\mathrm{ac}\ll I_\mathrm{th}$ }\\
      \left(2I_\mathrm{dc}+\sqrt{2}I_\mathrm{ac}\right)\sqrt{I_\mathrm{ac}^2-I_\mathrm{th}^2} & \text{$I_\mathrm{ac}\gg I_\mathrm{th}$. }
    \end{cases}
\end{eqnarray}

\subsection{Longitudinal modes: \label{LM}}

Here we estimate the differences in frequencies, and associated differences in applied magnetic field for measurements at constant frequency, of edge modes which would have different longitudinal wavelengths, in reference to the experimental results of Figure~\ref{Fig:VoltageSignals}(a). These estimates are based on the differences in frequencies of magnetostatic Damon-Eshbach surface modes \cite{DamEsh1960} of ferromagnetic films of thickness $2b$, whose direction of propagation is perpendicular to the applied magnetic field (as is the case of our Py stripe). In our notation, the frequencies of the Damon-Eshbach surface modes in the limit of small longitudinal wavevector $k$ are given by ($\omega=2\pi f$):
\begin{equation}
    f \simeq G[ \sqrt{h(h+1)}+kb/2\sqrt{h(h+1)} ] \; ,
\end{equation}
with $h=H_0/4\pi M_s$ representing the applied magnetic field (in our following estimates we take $H_0=500$\,Oe, corresponding to Figure~\ref{Fig:VoltageSignals}(a)), and $G=(|\gamma_{\text{Py}}|/2\pi) 4\pi M_s =21.7$\,GHz \cite{Duan2014b}. As discussed in the main text, due to pinning at the edges of the active region, the smallest wavevectors correspond to $k_j=2\pi/[3.6,1.8,1.2]\,\mu$m$^{-1}$, $j=1,2,3$. The corresponding frequencies are (except for the constant term $G\sqrt{h(h+1)}$) $f_1=0.18$\,GHz, $f_2=0.36$\,GHz and $f_3=0.54$\,GHz. Thus, the differences in frequencies of these longitudinal modes are $f_2-f_1=0.18$\,GHz\, $ =f_3-f_2$, at a fixed applied magnetic field. Approximating the slope of the experimental frequency vs. magnetic field of Fig. \ref{Fig:VoltageSpectra} as $\Delta f/\Delta H \simeq 3$\,GHz/500\,Oe, then the associated magnetic field differences between these modes (at fixed frequency as in Fig. \ref{Fig:VoltageSignals}) are $H_1-H_2 \simeq 30$\,Oe\,$\simeq H_2-H_3$.

\subsection{Oersted field calculation: \label{CO}}

If $I$ is the total current applied to the Py/Pt bilayer, then the current flowing in the Pt layer $I_\text{Pt}$ can be calculated using the parallel resistance model: $1/R=1/R_\text{Pt}+1/R_\text{Py}$, which gives:
\begin{equation}
I_\text{Pt} = I/(1+R_\text{Pt}/R_\text{Py}).
\end{equation}
Using the measured resistivity of Pt and Py films \cite{Duan2014}:  $\rho_\text{Pt} = 21.9$\,$\mu\Omega\,$cm and $\rho_\text{Py} = 65.2$\,$\mu\Omega$\,cm, we estimate $R_\text{Pt} \simeq R_\text{Py}/3$, i.e. $I_\text{Pt} \simeq (3/4) I$. The Oersted field in Py is generated by the current in Pt (the current in Py produces magnetic fields in Py that have null average over the Py layer thickness). We approximate the Oersted field applied to Py as due to an infinite sheet of current corresponding to the net current flowing through the Pt layer thickness in our experiment. In this approximation, Ampere's law (MKS units) gives:
\begin{equation}
H_{Oe} = j_\text{Pt} \Delta/2 = I_\text{Pt}/(2 w)\,\,\text{A\,m$^{-1}$} , 
\end{equation}
with $\Delta$ the thickness of Pt and $w$ the width of the nanowire (in Gaussian units $H_{Oe}=2\pi I_\text{Pt}/(10^3\,w)$ Oe, with $w=2c$). Then, the Oersted field due to Pt in Py is given by $H_{Oe}=I_\text{Pt}/(2w) = I_\text{Pt}(\text{A})/(2\times190\,\text{nm})=2.63 \times 10^3 I_\text{Pt}(\text{mA})\,\,\text{A\,m$^{-1}$}
=2.63 \times 10^3 (4\pi/10^3) I_\text{Pt}(\text{mA})\,\,\text{Oe} = 33\,I_\text{Pt}(\text{mA})$\,\,Oe. This leads to $h_{Oe}=H_{Oe}/4\pi M_s=\tilde{k}I_\text{Pt}=k I$, thus $k=33 (3/4)/(4\pi M_s) =3.24 \times 10^{-3}\,$mA$^{-1}$.

\subsection{Magnetization dynamics:}

The following terms contribute to the linear magnetization dynamics of Eq.\,(\ref{EqM1}):
\begin{eqnarray}
\frac{1}{m_s(x)} \frac{\delta U_Z}{\delta a^*} & = & h_x(\tau) a(x) , \label{duz} \\
\frac{1}{m_s(x)} \frac{\delta U_A}{\delta a^*}  & = & -k_s m_s(x)(a(x)-a^*(x)) ,  \\
\frac{1}{m_s(x)} \frac{\delta U_X}{\delta a^*}  & = & -\frac{d}{ m_s(x)} \nabla \cdot [ m_s^2(x) \nabla a(x) ] , \\
\frac{1}{m_s(x)}\frac{\delta U_D}{\delta a^*} & = &-  \int_{-1}^{1} dX'  m_s(X') (a(X')-a^*(X')) \nonumber \\
& & \times \ln \left(\frac{|X-X'|}{\sqrt{(X-X')^2+(2p)^2}}
\right)/4 \pi p  \nonumber \\
& &-H_V(X) a(x) ,
\label{dud} \\
\frac{1}{m_s(x)} \frac{\delta U_{STT}}{\delta a^*} & = & J m_s(x) a(x) ,
\label{dustt}
\end{eqnarray}
with $p=b/c$, and $\langle\vec{H}_D(M_s(x)\hatb{x})\rangle=-4\pi M_0 H_V(x)\hatb{x}$ ($\langle...\rangle$ means average over the thickness).

Now, the coefficients of the expansion Eq.\,(\ref{af}) for the dynamic variable $a(X,\tau)$, that satisfy the boundary condition $ a(X,\tau) =0$, at the edges are given by:
\begin{eqnarray}
a_l(\tau) & = & \frac2V \int dV \cos ((2l-1) \pi X/2) a(X,\tau) , \nonumber \\
f_l(\tau) & = & \frac2V \int dV \sin (l\pi X) a(X,\tau)  .
\label{abs}
\end{eqnarray}

According to Eqs.\,(\ref{EqM1}), (\ref{af}), (\ref{abs}), one has the following equations of motion for the time evolution of the coefficients $a_l(\tau), b_l(\tau)$:
\begin{eqnarray}
i \frac{d{a}_l }{d\tau} & = & \frac2V (1-i\alpha) \int dV  \frac{\cos ((2l-1)\pi X/2) }{m_s(X)} \frac{\delta U}{\delta a^*} , \label{eqal} \\
i \frac{d{f}_l}{d\tau} & = & \frac2V (1-i\alpha) \int dV  \frac{\sin (l\pi X)}{m_s(X)} \frac{\delta U}{\delta a^*} . \label{eqbl}
\end{eqnarray}
Due to symmetry considerations the previous equations separate, i.e. $\dot{a}_l$ depends only on $a_j$'s and $a_i^*$'s, and similarly for $\dot{f}_l$, i.e. it depends
only on $f_j$'s and $f_i^*$'s. 

\subsubsection{Conservative equations of motion:}

In the conservative case Eq.\,(\ref{eqal}) for $\dot{a}_l$ becomes:

\begin{eqnarray}
i \dot{a}_l & = &  h_x(\tau) a_l -k_s \sum_n \tilde{A}_{ln}(a_n-a_n^*) \nonumber \\
& & + \frac{d}{c^2} \sum_n B_{ln} a_n \nonumber \\
& & - \sum_n \frac{C_{ln}}{4\pi p} (a_n-a_n^*) -\sum_n D_{ln} a_n ,
\label{alnd}
\end{eqnarray}
with 
\begin{eqnarray}
\tilde{A}_{ln} & = &  \int_{-1}^{1}dX m_s (X) \cos (k_l X) \cos (k_n X) , \label{aln} \\
B_{ln} & = &  k_n  \int_{-1}^{1}dX \frac{\cos (k_l X)}{m_s(X)} \frac{d}{dX} [m_s^2(X) \sin (k_n X) ] ,  \\
C_{ln} & = &  \int_{-1}^1 dX \int_{-1}^{1}dX' m_s (X') \cos (k_l X) \cos (k_n X')  \nonumber \\
 & & \times \ln \left( \frac{|X-X'|}{\sqrt{(X-X')^2+(2p)^2}}
 \right), \\
 D_{ln} & = &  \int_{-1}^{1}dX H_V(X) \cos (k_l X) \cos (k_n X) \label{dln} ,
\end{eqnarray}
with $k_l=(2l-1)\pi/2$, and similarly for $k_n$.

Expressions for these coefficients are given in the section \ref{coef} of this Appendix for the case in which dilution is assumed to occur linearly at a scale $L$ from each of the 
edges of the sample. 

If one looks for solutions of Eq.\,(\ref{alnd}) of the type:
\begin{equation}
a_l (\tau) = c_l \exp (-i \Omega \tau)+d_l \exp (i \Omega \tau)
\label{alti}
\end{equation}
i.e. 
\begin{equation}
a_l^* (\tau) = d_l^* \exp (-i \Omega \tau) +c_l^* \exp (i \Omega \tau) ,
\label{atpi}
\end{equation}
then the equations of motion (\ref{alnd}) lead to the eigenvalue problem $Mv = \Omega v$ (assuming $h_x$ independent of time), with 
the eigenvector $v^T=(c,d^*)=(c_n , d_n^*)$ (shorthand notation for an extended vector), and the matrix $M$ given as:
\begin{equation}
M = \left(
\begin{array}{cc}
R & S \\
-S^* & -R^*
\end{array}
\right)
=
\left(
\begin{array}{cc}
R & S \\
-S & -R
\end{array}
\right) ,
\end{equation}
with 
\begin{eqnarray}
R_{ln} & = &  h_x \delta_{ln} -k_s \tilde{A}_{ln} +\frac{d}{c^2} B_{ln} \nonumber \\
& & -D_{ln} - \frac{ C_{ln}}{4\pi p} ,   \label{rln} \\
S_{ln} & = & k_s \tilde{A}_{ln} +\frac{ C_{ln}}{4\pi p} . \label{sln}
\end{eqnarray}

\subsection{Linear dynamics including spin transfer torque and damping, dc current:}

In the presence of damping and spin transfer torque the equations of motion (\ref{eqal}) take the following form ($U=U_C+iU_{STT}$ is imaginary in this case):

\begin{equation}
i \dot{a}_l = (1-i \alpha) \sum_n [(R_{ln} +iJ \tilde{A}_{ln})a_n+S_{ln} a_n^*] \; ,
\label{emal}
\end{equation}
with $\tilde{A}_{ln},R_{ln}, S_{ln}$ the matrices given in Eqs.\,(\ref{aln},\ref{rln},\ref{sln}). Searching for solutions
of the type:
\begin{equation}
a_l (\tau) = c_l \exp (-i \Omega \tau -\nu \tau)+d_l \exp (i \Omega \tau -\nu \tau) 
\label{alt}
\end{equation}
i.e. 
\begin{equation}
a_l^* (\tau) = d_l^* \exp (-i \Omega \tau -\nu \tau) +c_l^* \exp (i \Omega \tau -\nu \tau) \; ,
\label{atn}
\end{equation}
 the equations of motion (\ref{emal})
and their complex conjugates, become the eigenvalue problem $\tilde{M} v = \tilde{\delta} v$, with 
\begin{equation}
\tilde{M} = \left(
\begin{array}{cc}
(1-i\alpha) \tilde{R} & (1-i\alpha) S \\
-(1+i \alpha) S^* & -(1+i \alpha) \tilde{R}^*
\end{array}
\right) \; , \label{Mt}
\end{equation}
with $\tilde{R}=R+iJ \tilde{A}$, $\tilde{\delta} = \Omega -i \nu$, and $v^T=(c^T,(d^*)^T)$.
The eigenmodes of this problem that includes damping and spin transfer torque, may be found by finding the right eigenvectors of $\tilde{M}$. These eigenvectors will be the columns of a matrix $W$ that defines a change of variables to the amplitudes of the eigenmodes $b_l, b_l^*$, as follows:
\begin{equation}
\left( 
\begin{array}{c}
a \\ a^*
\end{array}
\right)
= W \left( 
\begin{array}{c}
b \\ b^*
\end{array}
\right) .
\end{equation}
The equations of motion (\ref{emal}) (and their complex conjugates) may be written as:
\begin{equation}
i \left(
\begin{array}{c}
\dot{a} \\ \dot{a}^*
\end{array}
\right)
= 
\tilde{M}
\left(
\begin{array}{c}
{a} \\ {a}^*
\end{array}
\right) .
\end{equation}
Multiplying this equation on the left by $W^{-1}$ (the left eigenvectors of $\tilde{M}$) one gets 
the diagonal equation of motion for the amplitudes of the eigenmodes:

\begin{equation}
i \left(
\begin{array}{c}
\dot{b} \\ \dot{b}^*
\end{array}
\right)
= 
\tilde{D}
\left(
\begin{array}{c}
{b} \\ {b}^*
\end{array}
\right) ,
\label{emdsn}
\end{equation}
with $\tilde{D}=W^{-1} \tilde{M} W$ a diagonal matrix, whose elements are the frequencies of the modes with associated imaginary parts as decay/growth rates. 

\subsection{Dipolar energy of a transversely magnetized stripe:}

The scaled dipolar energy is given by:
\begin{eqnarray}
U_D & = & -\frac{1}{8\pi M_0^2} \int dV \vec{H}_D(\vec{M}) \cdot \vec{M} \nonumber \\
& =  & -\frac{1}{8\pi M_0^2} \int dV \langle \vec{H}_D(\vec{M}) \rangle \cdot \vec{M} \; , 
\end{eqnarray}
where $\langle... \rangle$ is average over the thickness: the second equality follows since in our model the magnetization does not vary over the thickness. Now
$\vec{H}_D(M_z \hatb{z})=0$ since $M_z \hatb{z}$ does not have surface or volume charges associated. According to Ref.~\cite{Duan2015} ($p = b/c$):

\begin{eqnarray}
& & \langle\vec{H}_D(M_y \hatb{y})\rangle_y(X)= \nonumber \\ &&
-\frac{1}{p}
\int_{-1}^1 dX' M_y(X') \ln(1+(2p/(X-X'))^2) \: ,
\end{eqnarray}
with $X \equiv x/c$. Also,
\begin{eqnarray}
\langle\vec{H}_D(M_x \hatb{x})\rangle & =& \langle\vec{H}_D(M_s(X) \hatb{x})\rangle
\nonumber \\
- \langle\vec{H}_D(M_s(X) aa^* \hatb{x})\rangle \; ,
\end{eqnarray}
with $\langle\vec{H}_D(M_s(X) \hatb{x})\rangle(X) \equiv -4\pi M_0 H_V(X) \hatb{x}$, and only due to magnetic volume charges (it is assumed that at the edges of the stripe the magnetization goes to zero).

Using the reciprocity theorem ($\int_V \vec{m}_1 \cdot \vec{H}_D(\vec{m}_2) = \int_V \vec{m}_2 \cdot \vec{H}_D(\vec{m}_1)$ for any two magnetization configurations), 
and using the nonzero components of the average demagnetizing field, one obtains the following expression for the demagnetizing energy:
\begin{eqnarray}
U_D & = & -\frac{V}{8\pi p} \int_{-1}^{1} dX  \int_{-1}^{1} dX' m_s(X) m_s(X')  \nonumber \\
& & \times\, m_y(X) m_y(X') \ln (\frac{|X-X'|}{\sqrt{(X-X')^2+(2p)^2}} ) \nonumber \\
& & -\frac{V}{2} \int_{-1}^{1} dX H_V (X) m_s(X) aa^* \nonumber \\
& &  -\frac{1}{8\pi} \int dV m_s(X)\langle H_D^x(m_s(X)aa^* \hatb{x})\rangle aa^* \; . \nonumber \\
\label{UD}
\end{eqnarray}
Using that $m_y = -(i/2)(a-a^*)\sqrt{2-aa^*}$, to quadratic order in $a, a^*$ the previous expression for the demagnetizing energy is approximated as:
\begin{eqnarray}
U_D^{(2)} & = & \frac{ V}{16 \pi p} \int_{-1}^{1} dX  \int_{-1}^{1} dX' m_s(X)m_s(X')  \nonumber \\
& & \times\, (a(X)-a^*(X))(a(X')-a^*(X')) \nonumber \\
& & \times\,  \ln \left(\frac{|X-X'|}{\sqrt{(X-X')^2+(2p)^2}} \right) \nonumber \\
& &-\frac{V}{2} \int_{-1}^{1} dX H_V (X) m_s(X) a(X)a^*(X) \; ,
\label{ud2}
\end{eqnarray}
meaning that
\begin{eqnarray}
\frac{\delta U_D^{(2)}}{\delta a^*} & = &- \frac{m_s(X)}{4 \pi p}  \int_{-1}^{1} dX'  m_s(X') \nonumber \\
& & \times \, (a(X')-a^*(X')) \ln \left(\frac{|X-X'|}{\sqrt{(X-X')^2+(2p)^2}} \right) \nonumber \\
& &- H_V (X) m_s(X) a(X) \; .
\label{delud}
\end{eqnarray}

Going back to $H_V(x)$, to simplify the analysis we take first only the right edge region, and its contribution to $H_V(x) \hatb{x}$ would be given by (origin taken at the right edge (r), and $L$ is taken as the length of dilution):
\begin{eqnarray}
H_V^r(x) & = &  -\frac{1}{4\pi M_0 b} \int_{-L}^0 dx' 
(-\frac{\partial M_s(x')}{\partial x'} ) \nonumber \\ & & 
\times\, \int_{-b}^b dy \int_{-b}^b dy' \frac{(x-x')}{(y-y')^2+(x-x')^2}. \nonumber \\
\end{eqnarray}
The volume magnetic charge density at the right edge would be $(-M_s'(x))=\nu$, with $\nu=M_0/L$ a constant, then:
\begin{eqnarray}
H_V^r(X)  & = & -\frac{c \nu}{4\pi M_0} \int_0^{L/c} dX'  \nonumber \\ & & 
\times\, \int_{-1}^{1}dY \int_{-1}^{1}dY'
 \frac{(X+X')/p}{(Y-Y')^2+((X+X')/p)^2} \nonumber  \\
 & = & -\frac{b \nu}{8\pi M_0} \int_{-1}^{1}dY \int_{-1}^{1}dY'  \nonumber \\ & &
\times\, \ln [(Y-Y')^2+((X+X')/p)^2]_{0}^{L/c}. \nonumber \\
\end{eqnarray}
Introducing $q=L/b$, and with the change of variables $V=Y-Y'$ and $U=Y+Y'$, one obtains:
\begin{eqnarray}
H_V^r(X)  & = &-\frac{ 1}{4\pi q } \int_0^{2} dV (2-V) \nonumber \\  
&\times&\, [ \ln (V^2+((L/c+X)/p)^2)- \ln (V^2+(X/p)^2) ]. \nonumber \\
\end{eqnarray}
and
\begin{eqnarray}
& & \int_0^2 dV (2-V) \ln (V^2+w^2)  \nonumber \\
& = & w^2 \ln |w| -6 +(4-w^2)\ln \sqrt{4+w^2} +4w \tan^{-1}(2/w). \nonumber \\
\end{eqnarray}
Putting all this together in the experimental geometry, with an origin at the center of the stripe:
\begin{align}
 -  4\pi q & H_V(X) = (q+(X-1)/p)^2 \ln |q+(X-1)/p | \nonumber \\
& + (4-(q+\frac{(X-1)}{p})^2) \ln \sqrt{4+(q+\frac{(X-1)}{p})^2} \nonumber \\
& +  4 (q+(X-1)/p) \tan^{-1}(\frac{2}{q+(X-1)/p}) \nonumber \\
& -  ((X-1)/p)^2 \ln |(X-1)/p| \nonumber \\
& -  (4-((X-1)/p)^2) \ln \sqrt{4+((X-1)/p)^2} \nonumber \\
& -  4 ((X-1)/p) \tan^{-1}(\frac{2p}{(X-1)}) \nonumber \\
& + (q-(X+1)/p)^2 \ln |q-(X+1)/p | \nonumber \\
& + (4-(q-\frac{(X+1)}{p})^2) \ln \sqrt{4+(q-\frac{(X+1)}{p})^2} \nonumber \\
& +  4 (q-(X+1)/p) \tan^{-1}(\frac{2}{q-(X+1)/p}) \nonumber \\
& -  ((X+1)/p)^2 \ln |(X+1)/p| \nonumber \\
& -  (4-((X+1)/p)^2) \ln \sqrt{4+((X+1)/p)^2} \nonumber \\
& -  4 ((X+1)/p) \tan^{-1}(\frac{2p}{(X+1)}) .
\label{hvx}
\end{align}

\subsection{Coefficients of equations of motion: \label{coef}}

In this section we present in more detail the determination of the coefficients (\ref{aln}-\ref{dln}) appearing in the equations of motion (\ref{alnd}). Taking that the 
region of dilution occurs within a distance $L$ from each edge, and that it corresponds to a linear growth of the material from the edge, we define $r=(c-L)/c$. Also $k_l=(2l-1)\pi/2$ and similarly for $k_n$. Then, 
\begin{eqnarray}
\hspace*{-5cm} \tilde{A}_{ln}  & = &\int_{-1}^{1}dX m_s (X) \cos (k_l X) \cos (k_n X) \nonumber \\
 &= & \frac{1}{(1-r)} \{  \frac{\cos ((k_l+k_n) r)-\cos((k_l+k_n)}{(k_l+k_n)^2}  \nonumber \\
&& + \frac{\cos ((k_l-k_n) r)-\cos(k_l-k_n)}{(k_l-k_n)^2} 
\},
\end{eqnarray}
\begin{eqnarray}
\tilde{A}_{nn} & = & \frac{1}{4} \{ 2(1+r) +\frac{1}{(1-r)k_n^2}[\cos (2 k_n r)-\cos(2k_n)] \}, \nonumber \\
\end{eqnarray}
\begin{eqnarray}
B_{ln} & = &  k_n  \int_{-1}^{1}dX \frac{\cos (k_l X)}{m_s(X)} \frac{d}{dX} [m_s^2(X) \sin (k_n X) ]  \nonumber \\
 & = & k_n^2 A_{ln} \nonumber \\
& & - \frac{2k_n}{(1-r)} \{  \frac{\cos ((k_l+k_n) r)-\cos(k_l+k_n)}{(k_l+k_n)}  \nonumber \\
&  &- \frac{\cos ((k_l-k_n) r)-\cos(k_l-k_n)}{(k_l-k_n)} \} ,
\end{eqnarray}
\begin{eqnarray}
B_{nn} & = & A_{nn} k_n^2+ \frac{\cos (2k_n)-\cos (2k_n r)}{(1-r)} ,
\end{eqnarray}
\begin{eqnarray}
C_{ln} & = & 
  \int_{-1}^1 dX \int_{-1}^{1}dX' m_s (X') \cos (k_l X)  \nonumber \\
 & &\times\, \cos (k_n X')\ln ( \frac{|X-X'|}{\sqrt{(X-X')^2+(2p)^2}}) \nonumber \\
 & = & 2 \int_{-1}^1 dX \int_{0}^{1}dX' m_s (X') \cos (k_l  X) \nonumber \\
 & &\times\, \cos (k_n X') \ln ( \frac{|X-X'|}{\sqrt{(X-X')^2+(2p)^2}}) \nonumber \\
 & = &  \int dV \int dU m_s((U-V)/2)  \ln ( \frac{|V|}{\sqrt{V^2+(2p)^2}}) \nonumber \\
 & & \times\, \{ \cos (k_l U/2) \cos (k_n U/2) \cos (k_l V/2)\cos (k_n V/2) \nonumber \\
 & & +\cos (k_l U/2) \sin (k_n U/2) \cos (k_l V/2)\sin (k_n V/2) \nonumber \\
 & & - \sin (k_l U/2) \cos (k_n U/2) \sin (k_l V/2)\cos (k_n V/2) \nonumber \\
 & & - \sin (k_l U/2) \sin (k_n U/2) \sin (k_l V/2)\sin (k_n V/2) \},\nonumber \\
 \label{cln}
 \end{eqnarray}
 where 
\begin{equation}
\begin{array}{ccc}
 U = X+X'  & ,  & V=X-X' , \\
 X=(U+V)/2 & , & X' = (U-V)/2,
 \end{array}
 \end{equation}
 
 \begin{equation}
  m_s(X')= 
  \left\{
\begin{array}{ccc}
1 & : & 0 \leq X'=\frac{U-V}{2} \leq r \\
\frac{1-X'}{1-r}=\frac{2+V-U}{2(1-r)} & : & r \leq X' \leq 1 .
\end{array}
\right .
 \end{equation}
 
Also, 
\begin{align}
2 \int_{-1}^1 dX & \int_{0}^{1}dX' \nonumber \\
& =  \int_{-2}^{-1-r} dV \int_{-V-2}^{V+2} dU+\int_{-1-r}^{0} dV \int_{V+2r}^{V+2} dU \nonumber \\ 
& + \int_{0}^{1-r} dV \int_{V+2r}^{-V+2} dU +\int_{-1-r}^{-1} dV \int_{-V-2}^{V+2r} dU \nonumber \\
& + \int_{-1}^{1-r} dV \int_{V}^{V+2r} dU +\int_{1-r}^{1} dV \int_{V}^{-V+2} dU \nonumber \\ ,
\end{align}
which has been separated according to the regions where $m_s(X')$ is not equal to one (first three), or equal to one (second one). 
 
Also:
\begin{eqnarray}
 & & \int dU \cos (l\pi U/2) \cos (n\pi U/2) = \nonumber \\
 & & \frac{\sin ((k_l+k_n) U/2)}{(k_l+k_n)} +\frac{\sin ((k_l-k_n) U/2)}{(k_l-k_n)} , \nonumber \\
 & &  \int dU \sin (k_l U/2) \cos (k_n U/2) =  \nonumber \\
  & & -\frac{\cos ((k_l+k_n) U/2)}{(k_l+k_n)} -\frac{\cos ((k_l-k_n)U/2)}{(k_l-k_n)} , \nonumber \\
 & &  \int dU \sin (k_l U/2) \sin (k_n U/2) =  \nonumber \\
  & &  -\frac{\sin ((k_l+k_n) U/2)}{(k_l+k_n)} +\frac{\sin ((k_l-k_n) U/2)}{(k_l-k_n)} , \nonumber \\
 & &  \int dU U \cos (k_l U/2) \cos (k_n U/2) =  \nonumber \\
 & & U\{ \frac{\sin ((k_l+k_n) U/2)}{(k_l+k_n)} 
 +\frac{\sin ((k_l-k_n) U/2)}{(k_l-k_n)} \} \nonumber \\
 & & +2\frac{\cos ((k_l+k_n) U/2)}{(k_l+k_n)^2} 
 +2\frac{\cos ((k_l-k_n) U/2)}{(k_l-k_n)^2} , \nonumber \\
 & &  \int dU U \sin (k_l U/2) \cos (k_n U/2) =  \nonumber \\
 & & -U\{ \frac{\cos ((k_l+k_n) U/2)}{(k_l+k_n)} 
 +\frac{\cos ((k_l-k_n) U/2)}{(k_l-k_n)} \} \nonumber \\
 & & +2\frac{\sin ((k_l+k_n) U/2)}{(k_l+k_n)^2} 
 +2\frac{\sin ((k_l-k_n) U/2)}{(k_l-k_n)^2} , \nonumber \\
 & &  \int dU U \sin (k_l U/2) \sin (k_n U/2) =  \nonumber \\
 & & U\{ -\frac{\sin ((k_l+k_n) U/2)}{(k_l+k_n)} 
 +\frac{\sin ((k_l-k_n) U/2)}{(k_l-k_n) } \} \nonumber \\
 & & -2\frac{\cos ((k_l+k_n) U/2)}{(k_l+k_n)^2} 
 +2\frac{\cos ((k_l-k_n) U/2)}{(k_l-k_n)^2} . \nonumber \\
\label{inu}
\end{eqnarray}
 
From these equations (\ref{inu}) one deduces:
\begin{eqnarray}
 & & \int dU \{ \nonumber \\
 & & \cos (k_l U/2) \cos (k_n U/2) \cos (k_l V/2)\cos (k_n V/2) \nonumber \\
 & & +\cos (k_l U/2) \sin (k_n U/2) \cos (k_l V/2)\sin (k_n V/2) \nonumber \\
 & & - \sin (k_l U/2) \cos (k_n U/2) \sin (k_l V/2)\cos (k_n V/2) \nonumber \\
 & & - \sin (k_l U/2) \sin (k_n U/2) \sin (k_l V/2)\sin (k_n V/2) \} \; ,\nonumber \\
 & = & \sin ((k_l+k_n) U/2) \cos ((k_l-k_n) V/2)/(k_l+k_n)  \nonumber \\
 & & +\sin ((k_l-k_n) U/2) \cos ((k_l+k_n) V/2)/(k_l-k_n) \nonumber \\
 & & + \cos ((k_l+k_n) U/2) \sin ((k_l-k_n) V/2)/(k_l+k_n) \nonumber \\
 & & + \cos ((k_l-k_n) U/2) \sin ((k_l+k_n) V/2)/(k_l-k_n)  \} \nonumber \\
 & = &  \sin [(k_l+k_n) U/2 +(k_l-k_n) V/2] /(k_l+k_n)  \nonumber \\
 &  & + \sin [(k_l-k_n) U/2 +(k_l+k_n) V/2] /(k_l-k_n)  \nonumber \\
 & \equiv & au(U,V,kl,kn) \; .
\label{au}
\end{eqnarray}
Also, for $k_l=k_n=k$:
\begin{equation}
au(U,V,k,k) \equiv aue(U,V,k) = \frac{U}{2} \cos (kV) +\frac{\sin (kU)}{2k} \; .
\end{equation}

Similarly,
\begin{eqnarray}
 & &  \int dU U \{ \nonumber \\
 & & \cos (k_l U/2) \cos (k_n U/2) \cos (k_l V/2)\cos (k_n V/2) \nonumber \\
 & & +\cos (k_l U/2) \sin (k_n U/2) \cos (k_l V/2)\sin (k_n V/2) \nonumber \\
 & & - \sin (k_l U/2) \cos (k_n U/2) \sin (k_l V/2)\cos (k_n V/2) \nonumber \\
 & & - \sin (k_l U/2) \sin (k_n U/2) \sin (k_l V/2)\sin (k_n V/2) \} \nonumber \\
 & = & U au(U,V,kl,kn) \nonumber \\
 & & +2 \cos ((kl+kn)U/2) \cos ((kl-kn) V/2)/(kl+kn)^2  \nonumber \\
 & & +2 \cos ((kl-kn)U/2) \cos ((kl+kn) V/2)/(kl-kn)^2 \nonumber \\
 & & -2 \sin ((kl+kn)U/2) \sin ((kl-kn) V/2)/(kl+kn)^2 \nonumber \\
 & & -2 \sin ((kl-kn)U/2) \sin ((kl+kn) V/2)/(kl-kn)^2    \nonumber \\
 & = & U au(U,V,kl,kn) \nonumber \\
 & & +2 \cos [(kl+kn)U/2 +(kl-kn) V/2]/(kl+kn)^2  \nonumber \\
 & & +2 \cos [(kl-kn)U/2 +(kl+kn) V/2]/(kl-kn)^2 \nonumber \\
  & \equiv & bu(U,V,kl,kn) \; ,
\label{bu}
\end{eqnarray}
 
and for $k_l=k_n=k$:
\begin{eqnarray}
 bu(U,V,k,k) & \equiv & bue(U,V,k) = \frac{U^2}{4} \cos (kV) \nonumber \\
 & & +\frac{\cos (kU)}{2k^2}+\frac{U}{2k}\sin (kU) \; .
\end{eqnarray}

Now, we define:
 
\begin{eqnarray}
cd(U,V,kl,kn) & \equiv & \ln ( \frac{|V|}{\sqrt{V^2+(2p)^2}})au(U,V,kl,kn) , \nonumber \\ 
\label{cd} \\
cn(U,V,kl,kn) & \equiv & \frac{1}{2(1-r)}\ln ( \frac{|V|}{\sqrt{V^2+(2p)^2}})  \nonumber \\
\times\, [(2&+&V)au(U,V,kl,kn)-bu(U,V,kl,kn)] .
 \nonumber \\
 \label{cn}
 \end{eqnarray}
Then, 

\begin{eqnarray}
C_{ln} & = &  \frac{n_n n_l}{2} \{ \int_{-1-r}^{-1} dV cd(U,V,kl,kn)|_{-V-2}^{V+2r}  \nonumber \\
& & + \int_{-1}^{1-r} dV cd(U,V,kl,kn)|_{V}^{V+2r}  \nonumber \\
& & + \int_{1-r}^{1} dV cd(U,V,kl,kn)|_{V}^{-V+2} \nonumber \\
& & + \int_{-2}^{-1-r} dV cn(U,V,kl,kn)|_{-V-2}^{V+2}  \nonumber \\
& & + \int_{-1-r}^{0} dV cn(U,V,kl,kn)|_{V+2r}^{V+2}  \nonumber \\
& & + \int_{0}^{1-r} dV cn(U,V,kl,kn)|_{V+2r}^{-V+2} \}.
\end{eqnarray}

\subsection{Uniform mode, extended stripe or film limit: \label{tfl}}

In order to get analytic results in a simpler case, we develop the case of parametric resonance of a uniform
mode in an extended film (effects of the edges of the stripe neglected). 

The matrix $\tilde{M}$ in this case is the following (no ac current, $J_0$ comes from the dc current spin transfer torque,
$h_x$ includes a dc Oersted field, no anisotropy):
\begin{align}
&\tilde{M} = \nonumber \\
&\left(
\begin{array}{cc}
(1-i\alpha) (h_x+\frac12+i J_0) & -(1-i\alpha)/2 \\
(1+i \alpha)/2 & -(1+i \alpha)(h_x +\frac12 -i J_0)
\end{array}
\right) .
\end{align}

The change of variables to the amplitudes $b_0, b_0^*$ of the uniform eigenmode is as follows:

\begin{equation}
\left( 
\begin{array}{c}
a_0 \\ a_0^*
\end{array}
\right)
=
\left(
\begin{array}{cc}
\lambda & -\mu \\
-\mu^* & \lambda^*
\end{array}
\right)
\left( 
\begin{array}{c}
b_0 \\ b_0^*
\end{array}
\right) = W \left( 
\begin{array}{c}
b_0 \\ b_0^*
\end{array}
\right) .
\label{ab}
\end{equation}

The eigenvalues of $\tilde{M}$ are given by:
\begin{equation}
\gamma_{\pm} \simeq 
i (J_0-\alpha (h_x+1/2)) \pm \sqrt{(h_x+1+\alpha J_0)(h_x+\alpha J_0) } \; ,
\end{equation}
i.e. one identifies the critical value of $J_0$ as $J_0^c=\alpha (h_x+1/2)$, since the equation
of motion for $b_0, b_0^*$ are:
\begin{equation}
i \left( 
\begin{array}{c}
\dot{b}_0 \\ \dot{b}_0^*
\end{array}
\right) = \left(
\begin{array}{cc}
\gamma_+ & 0 \\
0 & \gamma_{-}
\end{array}
\right)
 \left( 
\begin{array}{c}
b_0 \\ b_0^*
\end{array}
\right) = D 
 \left( 
\begin{array}{c}
b_0 \\ b_0^*
\end{array}
\right)
\; ,
\end{equation}
thus 
$b_0=b_0^0 e^{-i \gamma_+ \tau}=b_0^0 e^{-i \Omega_0 \tau-\nu_0 \tau}$, with $\Omega_0 = \sqrt{(h_x+1+\alpha J_0)(h_x+\alpha J_0) }$, 
$\nu_0 = J_0^c-J_0$, and $b_0^*=b_0^{0*}e^{-i \gamma_- \tau}$. 
The eigenvectors of $\tilde{M}$ may be calculated (they are the columns of the matrix W in Eq.\,(\ref{ab})), and using the normalization
$|\lambda|^2-|\mu|^2=1$, they lead to: 
\begin{equation}
\begin{array}{ccc}
\mu = \frac{1}{(1+i \alpha)}\sqrt{\frac{A-\Omega_0}{2\Omega_0}} & , & \lambda=
-\sqrt{\frac{A+\Omega_0}{2\Omega_0}} \; ,
\end{array}
\end{equation}
with $A= h_x+1/2$. 
In this case $W^{-1}$ is given by:
\begin{equation}
W^{-1} = 
\left(
\begin{array}{cc}
\lambda^* & \mu \\
\mu^* & \lambda
\end{array}
\right) .
\end{equation}

The equation of motion with an ac current takes the form:

\begin{equation}
i \left( 
\begin{array}{c}
\dot{b}_0 \\ \dot{b}_0^*
\end{array}
\right) =  D 
 \left( 
\begin{array}{c}
b_0 \\ b_0^*
\end{array}
\right)
+W^{-1} \left(
\begin{array}{cc}
f(\tau) & 0 \\
0 & - f^*(\tau)
\end{array}
\right)
W
 \left( 
\begin{array}{c}
b_0 \\ b_0^*
\end{array}
\right)
\;
\end{equation}
with $f(\tau) = \cos (2 \Omega_p \tau) f_0$, and $f_0=(1-i \alpha)(-h_\mathrm{ac}+i J_\mathrm{ac})\sqrt{2}$. And 
\begin{eqnarray}
 & & W^{-1} \left(
\begin{array}{cc}
f_0 & 0 \\
0 & - f_0^*
\end{array}
\right)
W = \nonumber \\ & & 
\left(
\begin{array}{cc}
(|\lambda|^2 f_0 +f_0^* |\mu|^2) & -\lambda^* \mu (f_0+f_0^*) \\
\lambda \mu^* (f_0+f_0^*) & - ( |\lambda|^2 f_0^* +f_0 |\mu|^2)
\end{array}
\right) .
\nonumber \\
\end{eqnarray}
Considering only the resonant terms of the previous first equation, this equation becomes:
\begin{equation}
i \dot{b}_0 = 
(\Omega_0-i \nu_0) b_0 - \lambda^* \mu (f_0+f_0^*) e^{-i 2 \Omega_p \tau} b_0^*/2 \; , 
\end{equation}
with 
$- \lambda^* \mu (f_0+f_0^*) \simeq -h_\mathrm{ac}/\sqrt{2}\Omega_0$. Thus, looking for solutions of the type
$b_0 = b_0^0 \exp (-i \Omega_p \tau-\nu \tau)$, $b_0^* = b_0^{0*} \exp (i \Omega_p \tau-\nu \tau)$, one obtains the condition:
\begin{widetext}
\begin{eqnarray}
\left(
\begin{array}{cc}
(\Omega_p-\Omega_0)-i(\nu-\nu_0) & h_\mathrm{ac}/2\sqrt{2}\Omega_0 \\
h_\mathrm{ac}/2\sqrt{2}\Omega_0   & (\Omega_p-\Omega_0)+i(\nu-\nu_0)
\end{array}
\right)
\left(
\begin{array}{c}
b_0^0 \\
b_0^{0*}
\end{array}
\right) = 0 .
\end{eqnarray}
\end{widetext}
Thus, $N_{00*}=h_\mathrm{ac}/2\Omega_0=I_\mathrm{ac}
\hat{N}_{00*}$, i.e. $\hat{N}_{00*}$ is proportional to the Oersted field in this model (proportional to the real part of $f_0$ that does not depend on $\beta$). 
Imposing that the determinant of the previous equation to be zero leads to the condition:
\begin{equation}
(\nu-\nu_0)^2 = h_\mathrm{ac}^2/(2\sqrt{2}\Omega_0)^2 -(\Omega_p-\Omega_0)^2 .
\end{equation}
Thus, the lowest ac current for which a uniform  auto-oscillation occurs at a given dc current, corresponds to $\nu=0$, $\Omega_p=\Omega_0$, and leads 
to the threshold ac current condition:
\begin{equation}
\begin{array}{ccc}
|h_\mathrm{th}| = 2\sqrt{2} \Omega_0 \nu_0 & 
\leftrightarrow & |I_\mathrm{th}|= \frac{\sqrt{2}\nu_0}{|\hat{N}_{00*}|}
= 2\sqrt{2} \Omega_0 \nu_0/k ,
\end{array}
\label{cc}
\end{equation}
which is the equivalent threshold condition as in Eq.\,(\ref{ccn}) for a general mode $(n)$. 

\subsection{Spin Hall angle: \label{sha}}

In our notation the prefactor magnitude of the spin Hall torque is given by $|\gamma|4\pi J$ [see Eq.\,(\ref{LLG})]. We used $J=\beta I$, with $I$ the current through the bilayer. According to Ref.~\cite{Liu2011}, in our units:
\begin{equation}
    J=\frac{\hbar}{2ed_{\text{Py}}4\pi M_s^2}\frac{I_{\text{Pt}}}{d_{\text{Pt}}w} \theta_H = \beta I \; ,
\end{equation}
with $e$ the charge of the electron, $d_{\text{Py,Pt}}$ the thicknesses of Py and Pt, $w$ the width of the wire, and $I_{\text{Pt}} \simeq (3/4)I$. The latter expression allows to derive the spin Hall angle $\theta_H$ from $\beta$.

\bibliography{Citations}

\end{document}